\documentclass[11pt,a4paper]{article}

\usepackage{jheppub}
\usepackage[T1]{fontenc} 
\usepackage[utf8]{inputenc}
\usepackage{tikz}
\usepackage{tikz-cd}
\usepackage{ctable}

\title{\boldmath{The Classical Double Copy for M-theory from a Kerr-Schild Ansatz for Exceptional Field Theory}}

\author[a]{David S. Berman}
\author[b]{Kwangeon Kim}
\author[c,d]{Kanghoon Lee}

\affiliation[a]{Centre for Research in String Theory,\\School of Physics and Astronomy, Queen Mary University of London,London E1 4NS UK}
\affiliation[b]{Department of Physics, Yonsei University, Seoul 03722, Korea}
\affiliation[c]{Asia Pacific Center for Theoretical Physics, Postech, Pohang 37673, Korea}
\affiliation[d]{Department of Physics, Postech, Pohang 37673, Korea}

\emailAdd{d.s.berman@qmul.ac.uk}
\emailAdd{kim64656@yonsei.ac.kr}
\emailAdd{kanghoon.lee1@gmail.com}

\abstract{
We construct the classical double copy formalism for M-theory. This extends the current state of the art by including the three form potential of eleven dimensional supergravity along with the metric. The key for this extension is to construct a Kerr-Schild type  Ansatz for exceptional field theory. This Kerr-Schild Ansatz then allows us to find the solutions of charged objects such as the membrane from a set of single copy fields. The exceptional field theory formalism then automatically produces the IIB Kerr-Schild ansatz allowing the construction of the single copy for the fields of IIB supergravity (with manifest $SL(2)$ symmetry).
}

\begin{document}
\bibliographystyle{utphys}

\begin{flushright}
QMUL-PH-20-27	
\\
APCTP Pre2020 - 025
\end{flushright}
\maketitle

\section{Introduction}
The double copy was originally discovered in the context of the perturbative graviton scattering and provides a map between graviton and Yang-Mills amplitudes \cite{Kawai:1985xq,Bern:2008qj,Bern:2010ue,BjerrumBohr:2009rd,Stieberger:2009hq,Bern:2010yg,BjerrumBohr:2010zs,Feng:2010my,Tye:2010dd,Mafra:2011kj,Monteiro:2011pc,BjerrumBohr:2012mg}. Subsequent work showed how the double copy relationship applies for amplitudes of higher loops \cite{Bern:1998ug,Green:1982sw,Bern:1997nh,Carrasco:2011mn,Carrasco:2012ca,Mafra:2012kh,Boels:2013bi,Bjerrum-Bohr:2013iza,Bern:2013yya,Bern:2013qca,Nohle:2013bfa, Bern:2013uka,Naculich:2013xa,Du:2014uua,Mafra:2014gja,Bern:2014sna, Mafra:2015mja,He:2015wgf,Bern:2015ooa, Mogull:2015adi,Chiodaroli:2015rdg,Bern:2017ucb,Johansson:2015oia,Oxburgh:2012zr,White:2011yy,Melville:2013qca,Luna:2016idw,Saotome:2012vy,Vera:2012ds,Johansson:2013nsa,Johansson:2013aca}.

The natural question of whether the double copy could be non-perturbatively defined led to the construction of what has been called ``the classical double copy''. This is where exact classical gravitational solutions are mapped to solutions in Yang-Mills. \cite{Monteiro:2014cda, Luna:2015paa,Berman:2018hwd,Gurses:2018ckx,Sabharwal:2019ngs,Kim:2019jwm,Bahjat-Abbas:2020cyb,Alfonsi:2020lub,Keeler:2020rcv,Elor:2020nqe,Momeni:2020vvr,Alawadhi:2020jrv,Chacon:2020fmr}. (Non-exact classical solutions were explored in \cite{Anastasiou:2014qba,Borsten:2015pla,Anastasiou:2016csv,Anastasiou:2017nsz,Cardoso:2016ngt,Borsten:2017jpt,Anastasiou:2017taf,Anastasiou:2018rdx,LopesCardoso:2018xes}.) Following this, the S-duality of Yang-Mills acting on classical solutions was related to a set of solution generating transformations in gravity \cite{Huang:2019cja,Alawadhi:2019urr} which indicates the double copy goes beyond the perturbative theory. Other relevant recent work is in \cite{delaCruz:2020bbn,Casali:2020vuy,Moynihan:2020ejh,Godazgar:2020zbv,Emond:2020lwi}.

A central component in the construction of the classical double copy has been the use of the Kerr-Schild Ansatz. This ansatz dating from 1965, has been very useful historically in finding numerous interesting solutions to general relativity. The ansatz works by linearising Einstein equations and reducing them to Maxwell's equations and so provides a map between exact solutions of general relativity (with all its non-linearities) and solutions of Maxwell theory- that are of course linear. This has been extended in various ways, first in \cite{Ett:2010by} and then more recently in \cite{Cho:2019ype,Lee:2018gxc} where the NS two form of supergravity was included in the ansatz. The insight in \cite{Cho:2019ype,Lee:2018gxc} was to use the Double Field Theory \cite{Siegel:1993th,Siegel:1993bj,Hull:2009mi,Hohm:2010pp} description of supergravity, see \cite{Aldazabal:2013sca,Berman:2013eva,Hohm:2013bwa,Berman:2019biz} and references therein for a review. Double Field Theory combines the metric and two-form into a single object called ``the generalised metric" . The action and subsequent equations of motion are then written down in terms of this generalised metric. In \cite{Cho:2019ype,Lee:2018gxc} a Kerr-Schild type ansatz was constructed for the generalised metric. From this followed the construction of a classical double copy relation for the fields of the NS sector of supergravity ie. the metric and NS two form. The single copy theory is actually given by two Maxwell fields from which one can reconstruct the metric and two form. In some sense, the existence of a double copy incorporating the NS two form is no surprise. Early in the study of the origin of the perturbative double copy \cite{Kawai:1985xq} it was seen that one can think of the double copy as emerging from string theory where one relates the closed string amplitude to a product of two open string amplitudes. The closed string includes both the metric and two form, and the open string contains one Maxwell field thus from the string perspective the relation discovered in \cite{Lee:2018gxc} is no surprise.

What about M-theory or its low energy limit, eleven dimensional supergravity? Its bosonic degrees of freedom are a metric and three form potential, usually denoted $C_{(3)}$. It is very unclear from the perturbative string perspective whether one should expect a double copy relation to hold for M-theory. After all, the associated extended object for M-theory is the membrane whose spectrum certainly does not split into independent left and right movers as for the string and there is no notion of a closed membrane being thought of as a product of open membranes. Thus the usual perturbative string theory justifications for the double copy are absent. However, the role of double field theory indicates the contrary for the following heuristic reason. There is an M-theory extension of double field theory called ``exceptional field theory". Its development over many years \cite{Hull:2007zu,Pacheco:2008ps,Berman:2010is,Berman:2011jh,Coimbra:2011ky,Coimbra:2012af,Aldazabal:2013via,Aldazabal:2013mya,Hohm:2013vpa,Hohm:2013uia,Hohm:2014fxa,Abzalov:2015ega,Musaev:2015ces,Hohm:2015xna,Berman:2015rcc,Rosabal:2014rga,Cederwall:2015ica} showed how to reformulate eleven dimensional supergravity in terms of a generalised metric. For an extensive review of the area see \cite{Berman:2020tqn}. Over this period it was observed that {\it{what happens for double field theory, happens for exceptional field theory}} and one can show quite explicitly how they share similar structures.
Thus it is natural to seek a Kerr-Schild ansatz for exceptional field theory and in doing so linearise the exceptional field theory equations of motion and thus find a double copy formulation for the combination of metric and three form. Using exceptional field theory to do this will provide a bonus. As part of its formulation there is a choice to be made, called the ``choice of section condition''. Any exceptional field theory admits two distinct choices, one called the M-theory section, which  relates the generalised metric of the exceptional field theory to the fields of eleven dimensional supergravity. The other choice of section, ``the IIB section'', relates the generalised metric to IIB supergravity fields. Thus once we construct the Kerr-Schild Ansatz for the exceptional field theory we get for free both the M-theory and IIB double copies through choice of section.

This is what we will do in this paper. First we will construct a universal Kerr-Schild ansatz. This is an ansatz that will be valid for GR, DFT or Exceptional field theory that will reduce the theory to a set of linear equations. For the exceptional field theory we will take the $SL(5)$ theory though the structures are expected to be universal. We then show that solutions of the exceptional field theory with the Kerr-Schild Ansatz are given by a Maxwell field and a two form potential obeying their respective linear equations of motion. (In ExFT language, the single copy field of the $SL(5)$ exceptional field theory is a vector in the {\bf{10}} of $SL(5)$). We then describe the IIB section choice and derive a similar relation for the IIB theory. The resulting single copy for IIB then contains two Maxwell fields and an $SL(2)$ doublet which allows us to choose the $SL(2)$ frame of IIB.

\section{A Universal Kerr-Schild Ansatz}
In this section, we will consider the Kerr-Schild (KS) ansatz for General Relativity, Double Field Theory (DFT)  and Exceptional Field Theory (ExFT) in a single universal framework. The KS ansatz is exact without requiring any approximations and yet its form is similar to that of the ansatz used to calculate linear fluctuations. That is, it is written as a sum of two parts, a background metric that solves the equations of motion and something we call the fluctuation piece (though we don't require the fluctuations to be in any sense small and no subsequent approximations are made). Its crucial property is that once the (generalised) metric is placed in Kerr-Schild form, then the equations of motion for the fluctuation piece are reduced into a set of linear partial differential equations. Even though each theory has different properties (for example their actions are different), we are able to write a universal Kerr-Schild ansatz in terms of a null (generalised) vector and a projection operator which is defined for each theory. Note, that this does not mean that all solutions of general relativity or eleven dimensional supergravity are secretly linear. Far from it, the KS ansatz requires spacetimes to possess enough symmetry. Which spacetimes admit such a KS ansatz is discussed at length in the literature with the Petrov class providing a natural way to describe the necessary conditions (The equivalent of the Petrov classification has yet to be found for ExFT or DFT, it is hoped that this paper will stimulate work in that direction).

The presence of the projection operator is only needed for the ExFT or DFT cases and is trivial for GR. The reason for this projection operator in these cases is as follows. The generalised metric is a coset $G/H$ metric ($G$ and $H$ are listed in Table \ref{table:1}). The supergravity fields parametrise this coset and so the degrees of freedom of the supergravity fields match the number of degrees of freedom in the coset metric. Any arbitrary fluctuation of the generalised metric,$\delta \mathcal{M}_{MN}$ will not necessarily be consistent with this coset structure. To maintain the consistency of the theory we must only consider fluctuations $\delta \mathcal{M}_{MN}$ that preserve the coset structure of the metric (even off shell). This is done  by introducing a projection operator $P_{MN}{}^{PQ}$ so that perturbations of the generalised metric $\delta \mathcal{M}_{MN}$ satisfy \cite{Berkeley:2014nza,Berman:2019izh}
\begin{equation}
  \delta \mathcal{M}_{MN} = P_{MN}{}^{PQ} \delta \mathcal{M}_{PQ}\,,
\label{}\end{equation}
where $P_{MN}{}^{PQ}$ is the projection operator, projecting on to the adjoint representation
\begin{equation}
  P_{MN}{}^{KL} = \frac{1}{\alpha}\left(\delta_{M}^{(K} \delta_{N}^{L)}-\omega \mathcal{M}_{M N} \mathcal{M}^{K L}-\mathcal{M}_{M Q} Y^{Q(K}{}_{ R N} \mathcal{M}^{L) R}\right) \, .
\label{projection}\end{equation}
The constants $\alpha$ and $\omega$ depend on the theory and are listed in Table \ref{table:1}. 
\begin{table}[h!]
\centering
\begin{tabular}{ccccc} $G$ & $H$ & $H^{*}$ & $\alpha$ & $\omega$ \\ \hline $GL(d)$ & $SO(d)$ & $SO(1, d-1)$ & $1$ & 0 \\ $O(D, D)$ & $O(D) \times O(D)$ & $O(1, D-1) \times O(D-1,1)$ & $2$ & 0 \\ $SL(5)$ & $SO(5)$ & $SO(2,3)$ & 3 & $-\frac{1}{5}$ \\ $Spin(5,5)$ & $Spin(5) \times Spin(5)$ & $SO(5, \mathbb{C})$ & 4 & $-\frac{1}{4}$ \\ $E_{6(6)}$ & $USp(8)$ & $USp(4,4)$ & 6 & $-\frac{1}{3}$ \\ $E_{7(7)}$ & $SU(8)$ & $SU^{*}(8)$ & $12$ & $-\frac{1}{2}$ \\ $E_{8(8)}$ & $SO(16)$ & $SO^{*}(16)$ & $60$ & $-1$
\end{tabular}
\caption{List of duality groups $G$ and their maximal compact subgroups $H$ (Euclidean case) and $H^*$ (Lorentzian case)}
\label{table:1}\end{table}

We now present the Kerr-Schild ansatz for the generalised metric\footnote{The inverse of the generalized metric for DFT is identical with itself, $\mathcal{H}_{M}{}^{N} \mathcal{H}_{N}{}^{P} = \delta_{M}{}^{P}$, and the KS ansatz is given by $\mathcal{H}_{MN} = \mathcal{H}_{0MN} + \varphi P_{0MN}{}^{PQ} K_{P}K_{Q}$.}
\begin{equation}
\begin{aligned}
  \mathcal{M}_{MN} &= \mathcal{M}_{0 MN} +\kappa \varphi P_{0MN}{}^{PQ} K_{P} K_{Q}\,,
  \\
  \big(\mathcal{M}^{-1}\big){}^{MN} &= \big(\mathcal{M}^{-1}_{0}\big){}^{MN} - \kappa\varphi P_{0}{}^{MN}{}_{PQ} K^{P} K^{Q}\,.
\end{aligned}\label{Gen_KS_ansatz}
\end{equation}
$ \mathcal{M}_{0 MN}$ is the background generalised metric, this is often taken to be flat but it need not be. $K_{M}$ is a null vector with respect to $\mathcal{M}_{0}$, which explicitly means:\footnote{In DFT, the $\mathit{O}(D,D)$ metric $\mathcal{J}_{MN}$ defines the inner product, $\langle V, W\rangle = \mathcal{J}_{MN} V^{M} W^{N}$, instead of the generalised metric $\mathcal{H}_{MN}$. Thus, the null condition for the $\mathit{O}(D,D)$ case is replaced by $K_{M} \mathcal{J}^{MN} K_{N} = 0$. }
\begin{equation}
K_{M} \big(\mathcal{M}_{0}^{-1}\big){}^{MN} K_{N} = 0 \, .
\end{equation}
Finally, $\varphi$ is a scalar function. Here all the indices are raised and lowered by the background metric $\mathcal{M}_{0}$ (the  $\mathit{O}(D,D)$ metric $\mathcal{J}_{MN}$ for DFT case), and $P_{0MN}{}^{PQ}$ is the projection operator constructed with the background generalised metric. It turns out that the null condition on $K_{M}$ is not sufficient for the linear structure of the inverse metric. 
To make this so we need to impose an additional condition,
\begin{equation}
  Q_{MP} Q^{PQ} = 0\, .
\label{nil_Q}\end{equation}
Where
\begin{equation}
Q_{MN} = \varphi P_{0MN}{}^{PQ} K_{P} K_{Q} \,
\end{equation}
is the so called fluctuation piece. 
This condition provides a constraint on the maximal totally isotropic subspace $N$ (null space). For Riemannian geometry with Lorentzian signature $(-1,1,\cdots,1)$, the dimension of $N$, the so-called {\it{Witt index}}, is one. However, the Witt index of the generalised tangent space of ExFT or DFT is greater than one. This nilpotency condition then reduces the dimension of the null space. Since the structure of the generalised tangent space is different on a case by case basis, the Witt index of $N$ and the nilpotent conditions on $Q_{MN}$ are distinct as well. 

We expand $Q_{MN}$ more by using the definition of the projection operator \eqref{projection}, and it reduces to\footnote{See section \ref{sect_2.1} for DFT case.}
\begin{equation}
  Q^{M}{}_{N} = \frac{1}{\alpha} \Big(K^{M} K_{N} - Y^{MK}{}_{NL} K_{K} K^{L}\Big)\,.
\label{}\end{equation}
Then the nilpotency of $Q_{MN}$ implies $K_{M}$ satisfies the following relations, which is closely related to the closure of the generalised diffeomorphism in DFT and ExFTs,
\begin{equation}
\begin{aligned}
  Y^{M N}{}_{P Q} K_{M} K_{N}=0\,,
  \\
  Y^{M N}{}_{T Q} Y^{T P}{}_{R S} K_{N} K^{Q} K_{P} = 0\,.
\end{aligned}\label{reduced_Null}
\end{equation}
The first condition is the same as the section condition\footnote{We thank the anonymous referee for pointing this out.} 
\begin{equation}
  Y^{MN}{}_{PQ} \partial_{M} \otimes \partial_{N} = 0\,,
\label{}\end{equation}
and the second one comes from the nonlinear relation in \cite{Cederwall:2013naa}
\begin{equation}
  \left(Y^{M N}{}_{T Q} Y^{T P}{}_{R S} -Y^{M N}{}_{R S} \delta^{P}{}_{Q}\right) \partial_{(N} \otimes \partial_{P)}=0\,.
\label{}\end{equation}
Thus these conditions on the null vectors in \eqref{reduced_Null} defines the reduced null subspace. In the following we will review the known examples, GR and DFT, and introduce the KS ansatz for $SL(5)$ ExFT.

\subsection{The known examples: General Relativity and Double Field Theory}\label{sect_2.1}
We will briefly review the known cases from this perspective. The simplest example is GR since the projection operator for GR is trivial, 
\begin{equation}
  \big(P^{\scriptscriptstyle \mathrm{GR}}\big){}_{\mu\nu}{}^{\rho\sigma} = \delta_{\mu}{}^{(\rho} \delta_{\nu}{}^{\sigma)}\,.
\label{}\end{equation}
According to the universal form \eqref{Gen_KS_ansatz}, the KS ansatz for GR is given by 
\begin{equation}
\begin{aligned}
    g_{\mu\nu} &= \tilde{g}_{\mu\nu} + \kappa \varphi K_{\mu} K_{\nu}\,,
    \\
    \big(g^{-1}\big){}^{\mu\nu} &= \big(\tilde{g}^{-1}\big){}^{\mu\nu} - \kappa \varphi K^{\mu} K^{\nu}\,,
\end{aligned}\label{}
\end{equation}
where $\tilde{g}$ is a background metric and the vector field $K^{\mu}$ is null with respect to the background metric, that is it obeys:
\begin{equation}
K^\mu \tilde{g}_{\mu\nu} K^\nu =0 \, .
\end{equation}
One can then show that the additional nilpotency condition on $Q$ is trivial, because the projection operator is trivial. 
With this ansatz for the metric, the Einstein equations reduce to a linear equation for $\varphi K_{\mu}$ which maybe be identified as the single copy Maxwell field.

We now review the KS ansatz for DFT. The generalised metric $\mathcal{H}_{MN}$ on the $2D$ dimensional doubled space is given by the coset $\mathit{O}(D,D)/\mathit{O}(1,D-1)\times \mathit{O}(1,D-1)$. In terms of the usual $D$ dimensional metric $g$ and the Kalb-Ramond two-form $B$ it is written as follows:
\begin{equation}
  \mathcal{H}_{MN}=\begin{pmatrix}  g_{\mu\nu} -  B_{\mu\rho} g^{\rho\sigma} B_{\sigma\nu} & B_{\mu\rho}g^{\rho\nu} \\ -g^{\mu\rho}B_{\rho\nu} & g^{\mu\nu} \end{pmatrix}\, . \label{eq:genmetric}
\end{equation}
It is a symmetric $\mathit{O}(D,D)$ element and satisfies the $\mathit{O}(D,D)$ compatibility constraint, $\mathcal{H}_{MP} \mathcal{J}^{PQ} \mathcal{H}_{QN} = \mathcal{J}_{MN}$, where $\mathcal{J}_{MN}$ is the $\mathit{O}(D,D)$ metric which may be chosen to be:
\begin{equation}
  \mathcal{J}_{MN} = \begin{pmatrix} \mathbf{0} & \delta^{a}{}_{b} \\ \delta_{a}{}^{b} & \mathbf{0} \end{pmatrix}\,.
\label{}\end{equation}
In DFT the inner product for the $\mathit{O}(D,D)$ generalised tanget space is given by $\mathcal{J}$ instead of the generalised metric $\mathcal{H}$. However, the overall structure of the KS ansatz in DFT is the same with ExFTs.

The projector uses the so called $Y$-tensor which is given in terms of the $\mathit{O}(D,D)$ metric as follows
\begin{equation}
  Y^{PQ}{}_{MN}= \mathcal{J}^{PQ} \mathcal{J}_{MN}\, .
\label{}\end{equation}
From Table \ref{table:1}, we can read off the projection operator.  Remarkably for DFT it factorises as follows:
\begin{equation}
  \big(P^{\scriptscriptstyle \mathrm{DFT}}\big){}_{MN}{}^{PQ} = 2 P_{(M}{}^{P} \bar{P}_{N)}{}^{Q}\,,
\label{}\end{equation}
where $P_{M}{}^{N}$ and $\bar{P}_{M}{}^{N}$ are projectors defined by the generalized metric of DFT, $\mathcal{H}_{MN}$, and the $O(D,D)$ metric  $\mathcal{J}_{MN}$:
\begin{equation}
  P_{MN} = \frac{1}{2}\big(\mathcal{J}_{MN} + \mathcal{H}_{MN}\big) \,, \qquad \bar{P}_{MN} =\frac{1}{2}\big(\mathcal{J}_{MN} - \mathcal{H}_{MN}\big)\,.
\label{}\end{equation}
The factorisation of the projector, should not be a surprise, it is in fact a consequence of the left and right decomposition of the closed string modes.

Applying to the general KS ansatz \eqref{Gen_KS_ansatz}, we have the KS ansatz for $\mathcal{H}_{MN}$
\begin{equation}
\begin{aligned}
    \mathcal{H}_{MN} &= \mathcal{H}_{0MN} + \kappa \varphi \big(P^{\scriptscriptstyle \mathrm{DFT}}_{0}\big)_{MN}{}^{PQ} K_{P} K_{Q} \,,
\end{aligned}\label{DFT_KS}
\end{equation}
where $K_{M}$ is a null vector, $K_{M} \mathcal{J}^{MN} K_{N} = 0$ and $\big(P^{\scriptscriptstyle \mathrm{DFT}}_{0}\big)_{MN}{}^{PQ}$ is a background projection operator. It is useful to write the projected null vectors as
\begin{equation}
  L_{M} = P_{0M}{}^{N} K_{N}\,, \qquad \bar{L}_{M} = \bar{P}_{0M}{}^{N} K_{N}
\label{}\end{equation}
and from the completeness relation, $\delta_{M}{}^{N} = P_{M}{}^{N} + \bar{P}_{M}{}^{N}$, $K_{M}$ is decomposed to $L_{M}$ and $\bar{L}_{M}$
\begin{equation}
  K_{M} = L_{M} + \bar{L}_{M}\,.
\label{}\end{equation}
Then the finite fluctuation part is written as
\begin{equation}
  Q_{MN} = \varphi \big(L_{M}\bar{L}_{N} + L_{N}\bar{L}_{M}\big)\,.
\label{}\end{equation}

At this stage, $L$ and $\bar{L}$ do not have to be null. However, we need to require the nilpotency condition on $Q_{MN}$ \eqref{nil_Q} for the linearity of the KS ansatz,
\begin{equation}
  Q_{MN} Q^{NP} = \varphi^{2} L_{M} L^{P} \big(\bar{L}_{N}\bar{L}^{N}\big) + \varphi^{2} \bar{L}_{M} \bar{L}^{P} \big(L_{N} L^{N}\big) =0\,.
\label{}\end{equation}
This implies that $L_{M}$ and $\bar{L}_{M}$ have to be a pair of mutually orthogonal null vectors with definite chirality,
\begin{equation}
  L_{M} = P_{0M}{}^{N} L_{N} \,, \qquad \bar{L}_{M} = \bar{P}_{0M}{}^{N} \bar{L}_{N}
\label{chirality_null}\end{equation}
and
\begin{equation}
  L_{N} L^{N}=0\,, \qquad \bar{L}_{N}\bar{L}^{N} = 0\,.
\label{chirality_null2}\end{equation}
Thus the nilpotent condition of $Q_{MN}$ for DFT can be rephrased in terms of the chirality conditions \eqref{chirality_null}, and the Witt index for each chiral generalised tangent space is one \cite{Cho:2019ype,Lee:2015qza}. Furthermore, \eqref{chirality_null2} is consistent with the DFT section condition, $\mathcal{J}^{MN} \partial_{M} \otimes \partial_{N} = 0$, and the nonlinear relation \eqref{reduced_Null}. Then the KS ansatz is rewritten in a familiar form \cite{Lee:2018gxc}
\begin{equation}
    \mathcal{H}_{MN} = \mathcal{H}_{0MN} +\kappa \varphi \big(L_{M} \bar{L}_{N} + L_{N} \bar{L}_{M}\big)\,.
\label{}\end{equation}
The definite chirality is related to the left and right mover decomposition in closed string theory which plays a crucial role in double copy. How this generalises for M-theory and the fact that despite there not being a chiral decomposition there is still a nilpotent condition for $Q$ that allows the ansatz to work is one of the key results of this paper.

\subsection{Kerr-Schild ansatz for Exceptional Field Theory}
Exceptional field theory provides a way to combine the metric with the three form (and six form) potentials of eleven dimension supergravity into a single generalised metric. It does so in a way similar to DFT by extending the space in such a way that the generalised tangent bundle becomes augmented and the resulting generalised metric parametrises a coset. This coset is $E_d/H$ where H is the maximally compact subgroup of the exceptional group $E_d$. The space is then also equipped with an $E_d$ structure. Table 1, lists these cosets for different exceptional groups. (Note, that examining the Dynkin diagram will quickly reveal that for $d=5$ the exceptional group $E_5$ isomorphic to $Spin(5,5)$ and for $d=4$ it is isomorphic to $SL(5)$.) Unlike DFT it is not possible to write down the ExFT in arbitrary dimensions, although they share common structures. In each case one begins with eleven dimensional supergravity on a manifold, $M^{11}$ and then splits $M^{11}$ into $M^{d}\times M^{11-d}$. One then augments the $M^{d}$ part so that there is an associated exceptional $E_d$ geometry for that space. The specific way of constructing the exceptional geometry can be quite involved and is usually dealt with on a case by case basis. The reader is encouraged to read how this construction proceeds for all the different cases in  \cite{Berman:2020tqn} and references therein.

In this paper, to be concrete, from now on we will only consider the $SL(5)$ Exceptional Field Theory. Thus one carries out the following process. Start with 11-dimensional supergravity, then split it up into a four dimensional manifold (which in what follows we take to be Lorentzian) and a seven dimensional manifold. Even though this is an unusual convention, $E_{d(d)}$ duality structure still exists in the Lorentzian case \cite{Hull:1998br} and the associated ExFTs are constructed explicitly \cite{Blair:2013gqa,Berman:2019izh}. We will mostly ignore the seven dimensional manifold that goes along for the ride and is described by usual Riemannian geometry. The next step is to construct an exceptional generalised geometry by extending the four dimensional space so that its generalised tangent bundle $E$  includes both a vector field and a two-form field on the 4-dimensional base manifold $M_{4}$. These will then form a $\mathbf{10}$ representation of $SL(5)$. Thus the generalised tangent bundle for the $SL(5)$ theory will be
\begin{equation}
  E \simeq TM^{4} \oplus \Lambda^{2}T^{*}M^{4}\,,
\label{genTanSpace}\end{equation}
and the generalised tangent vector $V^{M}$  in the  $\mathbf{10}$ representation of $SL(5)$ is parametrised by a 4-vector $v^\mu$ and two-form ${\lambda_{\mu \nu}}$ in 4d as follows:
\begin{equation}
  V^{M} = \begin{pmatrix} v^{\mu} \\ \lambda_{\mu\nu}\end{pmatrix} \,,
\label{}\end{equation}
with $\mu, \nu(=0,1,2,3)$ being indices on the original $M^4$. On the other hand we may represent the $\mathbf{10}$ of $SL(5)$ ($M,N=1\cdots10$) as antisymmetric pairs of $\mathbf{5}$ indices, $[mn]$, where $m, n = 0,1,2,3 $ and $5$, so that
\begin{equation}
\begin{aligned}
   V^{[mn]} &= \begin{cases} V^{\mu 5} = -V^{5\mu} = v^{\mu} \\ V^{\mu\nu} = \tilde{\lambda}^{\mu\nu} = \frac{1}{2} \epsilon^{\mu\nu\rho\sigma} \lambda_{\rho\sigma}\end{cases}
\end{aligned}\label{}
\end{equation}
where $\epsilon_{\mu\nu\rho\sigma}$ is the Levi-Civita symbol, which is defined regardless of the metric signature, following the convention of \cite{Berman:2019izh}
\begin{equation}
  \epsilon_{0123} = 1\,, \qquad \epsilon^{0123} = 1\,.
\label{}\end{equation}
Note that we have put $\frac{1}{2}$ factor in a contraction of a pair of $\mathbf{5}$ indices like $V_{M} W^{M} = \frac{1}{2} V_{mm'} W^{mm'}$ to avoid overcounting. 

The $SL(5)$ generalized metric $\mathcal{M}_{MN}$ is then parametrized by the supergravity fields, $g_{\mu\nu}$ and $C_{\mu\nu\rho}$ as follows \cite{Berman:2010is}:
\begin{equation}
\begin{aligned}
  \mathcal{M}_{M N} &= |g|^{\frac{1}{5}}\begin{pmatrix}  g_{\mu \nu} +\frac{1}{4} C_{\mu \rho_{1} \rho_{2}} g^{\rho_{1} \rho_{2}, \sigma_{1} \sigma_{2}} C_{\sigma_{1} \sigma_{2} \nu} & ~~\frac{1}{2} C_{\mu \rho_{1} \rho_{2}} g^{\rho_{1} \rho_{2}, \nu_{1} \nu_{2}} \\ \frac{1}{2} g^{\mu_{1} \mu_{2}, \rho_{1} \rho_{2}} C_{\rho_{1} \rho_{2} \nu} &~~ g^{\mu_{1} \mu_{2}, \nu_{1} \nu_{2}}\end{pmatrix}\,,
  \\
  \left(\mathcal{M}^{-1}\right)^{M N} &= |g|^{-\frac{1}{5}} \begin{pmatrix} g^{\mu \nu} & -g^{\mu \rho} C_{\rho \nu_{1} \nu_{2}} \\ -C_{\mu_{1} \mu_{2} \rho} g^{\rho \nu} & ~~g_{\mu_{1} \mu_{2}, \nu_{1} \nu_{2}}+C_{\mu_{1} \mu_{2} \rho} g^{\rho \sigma} C_{\sigma \nu_{1} \nu_{2}}\end{pmatrix}\,,
\end{aligned}\label{para_genM10-1}
\end{equation}
or
\begin{equation}%
\begin{aligned}
  	\mathcal{M}_{m m',n n'} &= |g|^{\frac{1}{5}} \begin{pmatrix} g_{\mu \nu} + \frac{1}{4} C_{\mu\rho \rho'} g^{\rho\rho',\sigma\sigma'}C_{\sigma\sigma'\nu} &\frac{1}{4} C_{\mu\rho \rho'} g^{\rho\rho',\sigma\sigma'} \epsilon_{\sigma\sigma' \nu \nu'} \\ \frac{1}{4} \epsilon_{\mu \mu' \rho \rho'}g^{\rho\rho',\sigma\sigma'} C_{\sigma\sigma'\nu} & -|g|^{-1} g_{\mu_{1} \mu_{2}, \nu_{1} \nu_{2}} \end{pmatrix}\,,
  	\\
  	\left(\mathcal{M}^{-1}\right)^{mm',nn'} & =|g|^{-\frac{1}{5}} \begin{pmatrix} g^{\mu \nu} & -\frac{1}{2} g^{\mu \rho} C_{\rho\sigma\sigma'} \epsilon^{\sigma \sigma' \nu \nu'} \\ -\frac{1}{2} \epsilon^{\mu\mu' \rho\rho'} C_{\rho\rho' \sigma} g^{\sigma \nu} &-|g| g^{\mu \mu', \nu \nu'}+\frac{1}{4} \epsilon^{\mu \mu' \rho \rho'} C_{\rho \rho' \sigma} g^{\sigma \tau} C_{\tau \kappa \kappa'} \epsilon^{\kappa\kappa' \nu\nu'}\end{pmatrix}\,,
\end{aligned}\label{para_genM10-2}
\end{equation}
where $g_{\mu\mu',\nu\nu'}$ and $g^{\mu\mu',\nu\nu'}$ are defined in Appendix \ref{sec:C}. Note that $\mathcal{M}_{mm',nn',}$ can be represented by squaring the so-called little generalized metric $m_{mn}$ in $\mathbf{5}$-representation as \footnote{See Appendix \ref{sec:B} for the KS ansatz and equations of motion in $\mathbf{5}$-representation. }
\begin{equation}
  \mathcal{M}_{mm',nn'} = - \big(m_{mn} m_{m'n'} -m_{mn'} m_{m'n}\big)\,,
\label{little_metric}\end{equation}
where the overall minus sign in the right hand side is due to the Lorentzian metric signature. This is because to preserve the parametrization of the big generalised metric $\mathcal{M}_{mm',nn'}$ regardless of the metric signature, otherwise additional overall minus sign is required in the action depending on the choice of the metric signature (recall that action is odd power of $\mathcal{M}$). Thus there is a discrepancy for raising and lowering the $\mathbf{10}$-vector indices using the $\mathcal{M}_{mm',nn'}$ and $m_{mn}$. To avoid such confusions, we do not allow to use $m_{mn}$ for raising and lowering the $\mathbf{5}$-vector indices $m,n,\cdots$  within the $\mathbf{10}$-representation.

In the conventional $SL(5)$ ExFT, the manifold $M_{4}$ is considered to be a Euclidean internal space. On the contrary, here we assume that $M_{4}$ includes the time direction which allows the Lorentzian metric signature for defining null generalised vectors. Note that the Witt index for the Lorentzian $SL(5)$ generalised tangent space is four. In other words, there exist four mutually orthogonal null vectors out of ten-dimensional vector space. (In $\mathbf{5}$-representation, the Witt index is given by two) Recall that, the Witt index for Riemannian geometry with the Lorentzian metric signature is given by one, and there is no other null vector that is orthogonal to a given null vector. This feature yields a great power in manipulating null vectors and plays a crucial role to construct the linear equations of motion. Similarly, the Witt index for $\mathit{O}(D,D)$ generalised tangent space is $D$, however, the dimension of the null space reduces to one for each chiral null space when we impose the chirality on the null space as in \eqref{chirality_null}.

For simplicity we will now assume a flat background, $g_{\mu\nu} = \eta_{\mu\nu}$ and $C_{\mu\nu\rho} = 0$, so the background generalised metric $\mathcal{M}_{0}$ is:
\begin{equation}
  \mathcal{M}_{0MN} = \begin{pmatrix} \eta_{\mu\nu} &0 \\ 0 & \eta^{\mu\mu',\nu\nu'} \end{pmatrix} \,, \qquad  \mbox{and} \qquad \mathcal{M}_{0mm',nn'} = \begin{pmatrix} \eta_{\mu\nu} &0 \\ 0 & - \eta_{\mu\mu',\nu\nu'} \end{pmatrix} \,,
\label{flat_gen_metric}\end{equation}
and the associated null vector $K_{M}$ is parametrized as
\begin{equation}
  K_{M} = \begin{pmatrix} l_{\mu} \\ k^{\nu\nu'} \end{pmatrix}\,, \qquad K^{M} = \begin{pmatrix} l^{\mu} \\ k_{\nu\nu'} \end{pmatrix}\,,
\label{}\end{equation}
and
\begin{equation}
  K_{mm'} = \begin{cases}
  K_{\mu5} = l_{\mu}
  \\
  K_{\nu\nu'} = \tilde{k}_{\nu\nu'}
  \end{cases}
  \qquad 
  K^{mm'} = \begin{cases}
  K^{\mu5} = l^{\mu}
  \\
  K^{\nu\nu'} = - \tilde{k}^{\nu\nu'}
  \end{cases}\,.
\label{comp_null}\end{equation}
Here the Greek indices are raised and lowered by the flat background metric $\eta_{\mu\nu}$. From \eqref{Levi-Civita} $\tilde{k}_{\mu\nu}$ and $\tilde{k}^{\mu\nu}$ are related with $k_{\mu\mu'}$ and $k^{\mu\mu'}$ as
\begin{equation}
  \tilde{k}_{\mu\mu'} = \frac{1}{2} \epsilon_{\mu\mu'\nu\nu'} k^{\nu\nu'}\,, \qquad \tilde{k}^{\mu\mu'} = \frac{1}{2}\eta^{\mu\mu',\nu\nu'} \tilde{k}_{\nu\nu'}  = -\frac{1}{2} \epsilon^{\mu\mu'\nu\nu'} k_{\nu\nu'}\,,
\label{}\end{equation}
and their inner product satisfies
\begin{equation}
  k_{\mu\mu'} k^{\mu\mu'} = - \tilde{k}_{\mu\mu'} \tilde{k}^{\mu\mu'}\,.
\label{}\end{equation}
Then the null condition for $K$ is then
\begin{equation}
  K_{M} K^{M} = K_{mm'} K^{mm'}= l_{\mu} l^{\mu} + \frac{1}{2} k_{\nu\nu'}k^{\nu\nu'} = l_{\mu} l^{\mu} - \frac{1}{2} \tilde{k}_{\nu\nu'}\tilde{k}^{\nu\nu'} = 0\,.
\label{}\end{equation}
 The necessary projection operator $P_{M N}{}^{P Q}$ for the $SL(5)$ ExFT is given by:
\begin{equation}
  P_{M N}{}^{{P Q}} = \frac{1}{3}\Big(\delta_{(M}{}^{P} \delta_{N)}{}^{Q} +\frac{1}{5} \mathcal{M}_{MN} \mathcal{M}^{P Q}-\mathcal{M}_{M R} Y^{R(P}{}_{S N} \mathcal{M}_{}{}^{Q) S}\Big)\,,
\label{Projection_SL5}\end{equation}
and the $Y$-tensor is
\begin{equation}
  Y^{MN}{}_{PQ} = Y^{mm'nn'}{}_{pp'qq'} = \epsilon^{mm'nn'r} \epsilon_{pp'qq'r} \, .
\label{}\end{equation}
Combining the above results allows us to write the following Kerr-Schild ansatz for $\mathcal{M}_{MN}$
\begin{equation}
\begin{aligned}
  \mathcal{M}_{MN} &= \mathcal{M}_{0MN} +\kappa\varphi P_{0 M N}{}^{P Q}K_{P}K{}_{Q} \,,
  \\
  (\mathcal{M}^{-1})^{MN}  &= \mathcal{M}_{0}{}^{MN} - \kappa\varphi P_{0}{}^{M N}{}_{P Q}K^{P}K^{Q} \,,
\end{aligned}\label{KS_SL5}
\end{equation}
where $\kappa$ is some constant parameter and the indices are raised and lowered by the background generalized metric $\mathcal{M}_{0MN}$ and $(\mathcal{M}^{-1}_{0})^{MN}$ in \eqref{flat_gen_metric}. Let us denote the finite perturbation part of the KS ansatz as $Q_{mm'nn'}$. Then it can be evaluated explicitly
\begin{equation}
\begin{aligned}
  Q_{mm',nn'} &=\frac{1}{4} \varphi P_{0mm',nn'}{}^{pp',qq'} K_{pp'} K_{qq'} 
  \\
  & = 2\mathcal{M}_{0 mm',[n|p|}\Delta_{n']}{}^{p}
\end{aligned}\label{explicit_Q}
\end{equation}
where
\begin{equation}
  \Delta_{m}{}^{n} = \frac{1}{3}\varphi K_{mp} K^{np}\,.
\label{Delta}\end{equation}
By definition, $\Delta_{m}{}^{n}$ is traceless due to the null condition of $K_{M}$. We also need to further impose the nilpotency condition on $Q_{mm',nn'}$ so that the KS ansatz \eqref{KS_SL5} is exact. This requires
\begin{equation}
\begin{aligned}
  &\frac{1}{2} Q_{mm',pp'} Q^{pp',nn'} =0\,,
\end{aligned}\label{}
\end{equation}
which in turn implies that $\Delta_{m}{}^{n}$ is also nilpotent using \eqref{explicit_Q}
\begin{equation}
  \Delta_{m}{}^{n} \Delta_{n}{}^{p} = 0\,.
\label{}\end{equation}

One may check that \eqref{KS_SL5} is consistent with the relation to the small metric $m_{mn}$ in \eqref{little_metric} by the KS ansatz for $m_{mn}$ \eqref{gKS}
\begin{equation}
  m_{mn} = m_{0mn} + \Delta_{mn}\,, \qquad \qquad \Delta_{mn} = \kappa\varphi \ell_{m} \ell_{n}\,,
\label{}\end{equation}
where $\ell_{m}$ is a null vector in $\mathbf{5}$-representation. Thus we have the following relation between $K_{mm'}$ and $\ell_{m}$, 
\begin{equation}
  \frac{1}{3} K_{mp} K^{np} = \ell_{m} \ell^{n}\,.
\label{5d_10d_null}\end{equation}
This provides the relationship between the null condition on the {\bf{10}} space and the null condition in the {\bf{5}} space. It is then obvious that $\Delta_{m}{}^{n}$ satisfies the following identity
\begin{equation}
  \Delta_{m}{}^{[p} \Delta_{n}{}^{q]} = 0\,.
\label{DeltaSquqre}\end{equation}
Using the relation \eqref{5d_10d_null}, it is useful to represent $K_{mm'}$ using the null vectors $\ell_{m}$ and an auxiliary vector $j_{m}$ up to $SL(5)$ rotation
\begin{equation}
  K_{mm'} = \ell_{[m} j_{m']}\,,\qquad K^{mm'} = - \ell^{[m} j^{m']}\,,
\label{}\end{equation}
where $j_{m}$ is orthogonal to $\ell_{m}$ and normalised such that $j_{m} j^{m} = -12$. (Note, the minus sign on the right hand side of $K^{mm'}$ is due to the relation in \eqref{little_metric}. ) It is straightforward to show that $K_{mm'}$ satisfies the relations in \eqref{reduced_Null}. 

As we have seen before, the nilpotency of $Q_{mm',nn'}$ constrains the total null space $N$ into a reduced one, $\hat{N}$.  An arbitrary null vector $T_{mm'}$ resides in $\hat{N}$ satisfies 
%
\begin{equation}
\begin{aligned}
  &\frac{1}{3} T_{mp} T^{np} = \ell_{m}\ell^{n}\,\quad
  \mbox{or}
  &T_{mm'} = \ell_{[m} t_{m']}\,, 
\end{aligned}\label{null_vector}
\end{equation}
up to $SL(5)$ rotation. Here $t_{m}$ is a $\mathbf{5}$ vector  which is orthogonal to $\ell_{m}$, that is: $\ell_{m}\big(m_{0}^{-1}\big){}^{mn} t_{n} = 0$. It means any $\mathbf{10}$ null vector may be represented by a pair of $\mathbf{5}$ vectors, $\ell_{m}$ and $t_{m}$. 

Using the parametrization of $\mathcal{M}_{MN}$ \eqref{para_genM10-2}, we may then write the supergravity fields, $g_{\mu\nu}$ and $C_{\mu\nu\rho}$, in terms of the fields in the KS ansatz as follows,
\begin{equation}
\begin{aligned}
  |g| &= (1 -\kappa\Delta_{5}{}^{5})^{-\frac{5}{3}} 
  \\
  g^{\mu \nu} &= (1 - \kappa\Delta_{5}{}^{5})^{\frac{2}{3}} \Big(\eta^{\mu\nu}- \kappa\Delta^{\mu \nu}-\frac{\kappa^{2}}{1-\kappa \Delta_{5}{}^{5}} \Delta^{\mu 5} \Delta_{5}{}^{\nu} \Big) 
  \\
  g_{\mu \nu} &= (1 -\Delta_{5}{}^{5})^{-\frac{2}{3}}  \big(\eta_{\mu \nu} + \kappa\Delta_{\mu\nu} \big)  
  \\
  C_{\mu \nu \rho} &= \frac{\kappa}{1 -\kappa\Delta_{5}{}^{5}} \epsilon_{\mu \nu \rho \sigma} \Delta_{5}{}^{\sigma} \,.
\end{aligned}\label{KS_components}
\end{equation}
Here the flat background metric $\eta_{\mu\nu}$ is used for raising and lowering the Greek indices. One may note that the supergravity fields are all written in terms of $\Delta_{m}{}^{n}$, which is a contraction of two null vectors. This implies that there is a redundancy in the null vector $K_{mm'}$ for any given geometry compared to the $\mathbf{5}$-representation. The ten component $K_{mm'}$ ultimately has the same physical information as the $\mathbf{5}$ vector $l_{m}$. We may then rewrite \eqref{KS_components} in terms of the components of $K_{mm'}$
\begin{equation}
\begin{aligned}
  |g| &= \big|1 -\frac{\kappa \varphi}{3} l\cdot l\big|^{-\frac{5}{3}}  \,,
  \\
  g^{\mu \nu} &=\big|1 -\frac{\kappa \varphi}{3} l\cdot l\big|^{\frac{2}{3}} \Big(\eta^{\mu\nu}-\frac{\kappa\varphi}{3} \big(l^{\mu} l^{\nu} - \tilde{k}^{\mu\rho} \tilde{k}^{\nu}{}_{\rho}\big) -\frac{\kappa^{2}\varphi^{2}}{9(1 -\frac{\kappa \varphi}{3} l_{\mu} l^{\mu})} \tilde{k}^{\mu \rho} l_{\rho} \tilde{k}^{\nu \sigma} l_{\sigma}\Big) \,,
  \\
  g_{\mu \nu} &= \big|1 -\frac{\kappa \varphi}{3} l\cdot l\big|^{-\frac{2}{3}}  \Big(\eta_{\mu \nu} + \frac{\kappa\varphi}{3} \big(l_{\mu} l_{\nu} - \tilde{k}_{\mu\rho} \tilde{k}_{\nu}{}^{\rho}\big)\Big) \,,
  \\
  C_{\mu \nu \rho} &= \frac{2\kappa\varphi}{3|1 -\frac{\kappa \varphi}{3} l\cdot l|} l_{[\mu} k_{\nu \rho]}\,.
\end{aligned}\label{KS_components2}
\end{equation}
where $l\cdot l = l_{\mu} \eta^{\mu\nu} l_{\nu}$. 
Remarkably, the KS ansatz for the generalized metric is linear in $\kappa$, but component fields are highly nonlinear. If we set $k_{\mu\nu} = 0$, then $l^{\mu}$ becomes a null vector and the KS ansatz reduces to the conventional KS ansatz in GR, $g_{\mu\nu} = \eta_{\mu\nu} + \kappa\varphi l_{\mu} l_{\nu}$ and $C_{\mu\nu\rho} = 0$.

 Based on these results, we will show in the next section that the equations of motion of $SL(5)$ ExFT become linear.

\section{KS equations of motion}
The crucial feature of KS ansatz is the linearity of equations of motion. Under the KS ansatz, the Einstein equation can be reduced to linear equations, and many exact solutions are obtained by solving the linear equations of motion. The KS ansatz for DFT also produces linear equations of motion.
In this section, we use the KS ansatz for $SL(5)$ ExFT and show that the equations of motion are reduced to linear equations. 

The field content of the $SL(5)$ ExFT is the same as the 11-dimensional supergravity after the 11 = 7+4 decomposition with the tensor hierarchy fields,
\begin{equation}
  \big\{G_{ij} \,, A_{i}{}^{mm'}, B_{ijm},C_{ijk}{}^{m}, \mathcal{M}_{mm',nn'}\big\}\,,
\label{}\end{equation}
where $i,j,\cdots$ indices denote the 7-dimensional vector indices and $G_{ij}$ is the 7-dimensional metric. To match the $SL(5)$ ExFT with the 11-dimensional supergravity, we need the following Kaluza-Klein ansatz for the 11-dimensional metric $\hat{G}_{\hat{\mu}\hat{\nu}}$
\begin{equation}
  \hat{G}_{\hat{\mu}\hat{\nu}} = \begin{pmatrix} |\det g|^{-\frac{1}{5}} G_{ij} + A_{i}^{\mu}A_{j}{}^{\nu} g_{\mu\nu} & A_{i}{}^{\lambda} g_{\lambda\nu} \\ g_{\mu \lambda} A_{j}{}^{\lambda} & g_{\mu\nu} \end{pmatrix}\,.
\label{}\end{equation}

Here, we shall focus only on the (generalised) metric fields, $G_{ij}$ and $\mathcal{M}_{MN}$, and ignore all the form fields $A_{i},B_{ij}$ and $C_{ijk}$ for simplicity. We fix $G_{ij}$ and treat it as a non-dynamical field. In other words, we consider the equations of motion for $\mathcal{M}$ for a given $G_{ij}$. Of course, a restriction arises for $G_{ij}$ from the equations of motion of $\mathcal{M}$ as we will see later. Then the relevant $SL(5)$ ExFT action under the above assumptions is 
\begin{equation}
\begin{aligned}
  S &=\int_{\Sigma} \mathrm{d}^{7}z\, \mathrm{d}^{10}X \sqrt{|G|}~\Big[\ R[G_{ij}] + \frac{1}{12} G^{ij} \partial_{i} \mathcal{M}_{M N} \partial_{j} \mathcal{M}^{M N} -V[\mathcal{M},G]\ \Big]
  \\
  &=\int_{\Sigma} \mathrm{d}^{7}z\, \mathrm{d}^{10}X \sqrt{|G|}~\Big[\ R[G_{ij}] + \frac{1}{4} G^{ij}\Big( \partial_{i} m_{mn} \partial_{j} m^{mn} -\frac{1}{3} \partial_{i}\ln|m| \partial_{j}\ln|m| \Big)+V[m,G]\ \Big]
\end{aligned}\label{}
\end{equation}
where $R[G_{ij}]$ is the Ricci scalar with respect to $G_{ij}$, and the scalar potential $V[\mathcal{M},G]$ and $V[m,G]$ are given by
\begin{equation}
\begin{aligned}
 V[\mathcal{M},G] &= -\frac{1}{12} \mathcal{M}^{MN} \partial_{M} \mathcal{M}^{KL} \partial_{N}\mathcal{M}_{KL} + \frac{1}{2} \mathcal{M}^{MN} \partial_{M}\mathcal{M}^{KL} \partial_{K}\mathcal{M}_{LN} 
  \\
  &\quad -\frac{1}{2} \partial_{M} \mathcal{M}^{MN}\partial_{N} \ln |G| -\frac{1}{4} \mathcal{M}^{M N}\big(\partial_{M} G_{ij} \partial_{N} G^{ij} + (\partial_{M} \ln |G|) (\partial_{N} \ln |G|)\ \big) \,,
  \\
  V[m,G] &= -\frac{1}{8} m^{mp} m^{nq} \partial_{mn} m_{rs} \partial_{pq} m^{rs} -\frac{1}{2} m^{mp} m^{nq} \partial_{mn}m^{rs} \partial_{rp}m_{qs}  -\frac{1}{2} \partial_{mn} m^{mp} \partial_{pq}m^{nq} 
  \\
  &\quad -\frac{1}{2} m^{m p} \partial_{m n} m^{n q} \partial_{p q} \ln|G| -\frac{1}{8} m^{m p} m^{n q}\left(\partial_{m n} G^{ij} \partial_{p q} G_{ij}+\partial_{m n} \ln |G| \partial_{p q} \ln |G|\right)\,.
\end{aligned}\label{}
\end{equation}

The variation of the action with respect to the generalized metric $\mathcal{M}$ gives
\begin{equation}
  \delta_{\mathcal{M}} S = \int_{\Sigma} \mathrm{d}^{7}z\, \mathrm{d}^{10}X \sqrt{|G|}~ \delta\mathcal{M}^{MN} \mathcal{K}_{MN}\,,
\label{}\end{equation}
where
\begin{equation}
\begin{aligned}
  \mathcal{K}_{MN}&=- \frac{1}{6\sqrt{|G|}}\partial_{P} \big(\sqrt{|G|}\mathcal{M}^{PQ} \partial_{Q} \mathcal{M}_{MN}\big)+\frac{1}{\sqrt{|G|}} \partial_{P} \big(\sqrt{|G|}\mathcal{M}^{PQ} \partial_{(M} \mathcal{M}_{N)Q}\big) 
  \\
  &\quad +\frac{1}{12} \partial_{M} \mathcal{M}_{PQ} \partial_{N} \mathcal{M}^{PQ} 
   +\frac{1}{6} \mathcal{M}^{PQ} M^{RS} \partial_{P} \mathcal{M}_{R M} \partial_{Q} \mathcal{M}_{S N} 
  \\
  &\quad - \frac{1}{2} \mathcal{M}^{PQ} \mathcal{M}^{RS} \partial_{P} \mathcal{M}_{R M} \partial_{S} \mathcal{M}_{QN} - \frac{1}{2} \partial_{M}\partial_{N}\ln{|G|} +\frac{1}{4} \partial_{M} G_{ij} \partial_{N}G^{ij}
  \\
  &\quad - \frac{1}{6\sqrt{|G|}}\partial_{i} \big(\sqrt{|G|}G^{ij} \partial_{j} \mathcal{M}_{MN}\big)\,.
\end{aligned}\label{}
\end{equation}
As we have discussed in the previous section, we need the projection operator $P_{MN}{}^{PQ}$  \eqref{Projection_SL5} to get a consistent variation of $\mathcal{M}$ in order to have compatible with the variation of supergravity fields, $\delta g_{\mu\nu}$ and $\delta C_{\mu\nu\rho}$. Then the correct equations of motion is
\begin{equation}
  \hat{\mathcal{R}}_{MN} = P_{MN}{}^{PQ} \mathcal{K}_{PQ} = 0\,,
\label{eom1}\end{equation}
and it is the generalized curvature tensor for $SL(5)$ ExFT. Note that $\delta \mathcal{M}_{MN}$ is traceless because $\mathcal{M}\in SL(5)$, thus $\hat{\mathcal{R}}_{MN}$ have to be traceless as well. Here the hat marks the tracelessness. One may rewrite $\hat{R}_{MN}$ in terms of $\mathcal{R}_{MN}$ with the non-vanishing trace, $R_{M}{}^{M} \neq 0$,
\begin{equation}
  \hat{\mathcal{R}}_{MN} = \mathcal{R}_{MN} - \frac{1}{10} \mathcal{M}_{MN}(\mathcal{M}^{-1})^{PQ} \mathcal{R}_{PQ}\,,
\label{traceful_RMN}\end{equation}
From now on, we demand $\mathcal{R}_{MN}=0$ which is a stronger condition than $\hat{\mathcal{R}}_{MN}=0$, but it is enough for our purpose.

\subsection{Relating the equations of motion of $\mathcal{M}_{MN}$ and $m_{mn}$}
Using the relation between the generalised metric $\mathcal{M}_{mm',nn'}$ and the little metric $m_{mn}$  \eqref{little_metric}, we can identify the relation between the equations of motion for $\mathcal{M}_{mm',nn'}$ and $m_{mn}$. There is an advantage for using the small metric. Unlike $\delta \mathcal{M}_{mm',nn'}$, the variation of $m_{mn}$ does not require the projection operator. The only restriction on $\delta m$ is traceless condition, $\delta m_{mn} m^{mn} = 0$, and it can be implemented simply by using the trace part. Thus the structure of the equations of motion for $m_{mn}$, $\hat{\mathcal{R}}_{mn}$, is much simpler than the $\hat{\mathcal{R}}_{mm',nn'}$. 

The variation of $\mathcal{M}$ is related to $\delta m_{mn}$ as 
\begin{equation}
\begin{aligned}
  \delta \mathcal{M}_{mm',nn'} 
  &=- \delta m_{m n} m_{m'n'}- m_{m n} \delta m_{m'n'} + \delta m_{m n'} m_{m'n} +m_{m n'} \delta m_{m'n}\,.
\end{aligned}\label{var_M_m}
\end{equation}
As a consistency check, one can show the following identity, which ensures that \eqref{var_M_m} is a consistent variation, using the fact that $\delta m_{mn}$ is traceless,
\begin{equation}
  \delta \mathcal{M}_{mm',nn'} = \frac{1}{4} P_{mm',nn'}{}^{pp',qq'} \delta\mathcal{M}_{pp',qq'}\,.
\label{}\end{equation}
This connection between $\delta \mathcal{M}_{mm',nn'}$ and $\delta m_{mn}$ enable us to construct the relation between the equations of motion for $\mathcal{M}_{MN}$ and $m_{mn}$, $\hat{\mathcal{R}}_{mm',nn'}$ and $\hat{\mathcal{R}}_{mn}$ respectively 
\begin{equation}
  \frac{1}{2} \hat{\mathcal{R}}_{m p, q r} (\mathcal{M}^{-1})^{n p, q r} = -m^{np}\hat{\mathcal{R}}_{mp} 
\label{}\end{equation}
or 	
\begin{equation}
  \hat{\mathcal{R}}_{m m',n n'} = - \frac{2}{3}\mathcal{M}_{ m m', p [n} \hat{\mathcal{R}}_{n']}{}^{p}\,.
\label{}\end{equation}

Just as we have introduced $\mathcal{R}_{MN}$ from the traceless generalized curvature in \eqref{traceful_RMN}, we introduce $R_{mn}$ which has non-vanishing trace from $\hat{\mathcal{R}}_{mn}$ as
\begin{equation}
  \hat{\mathcal{R}}_{mn} = \mathcal{R}_{mn} -\frac{1}{5} m_{mn} \mathcal{R}_{p}{}^{p}
\label{}\end{equation}
then $\mathcal{R}_{mm',nn'}$ and $\mathcal{R}_{mn}$ are related by
\begin{equation}
  \mathcal{R}_{m m', n n'} = - \frac{2}{3}\mathcal{M}_{ m m', p [n} {\mathcal{R}}_{n']}{}^{p}
\label{5d_10d_eom}\end{equation}
%

\subsection{KS equations of motion}
We now consider the equations of motion for the generalised metric, $\mathcal{R}_{MN} = 0$ under the KS ansatz \eqref{KS_SL5}. As in the cases of GR and DFT, our aim is to reduce $\mathcal{R}_{MN}$, which is a set of highly nonlinear differential equations, to linear partial differential equations. Recall that the KS ansatz alone is not sufficient for linearization of the equations of motion in GR and DFT. It is necessary to impose an additional constraint, such as the geodesic condition in GR. Since these constraints arise from the full contraction between the null vectors and their equations of motion, we shall call them as the on-shell constraints. In the same spirit, we may construct the on-shell constraint for $SL(5)$ ExFT by contracting the null $\mathbf{10}$-vector $K_{mm'}$ with $\mathcal{R}_{mm',nn'}$ as follows:
\begin{equation}
  \frac{1}{4} K^{mm'} K^{nn'} \mathcal{R}_{mm',nn'} = -\frac{1}{3} K_{p n} K^{nn'} \mathcal{R}^{p}{}_{n'} = \ell^{m} \ell^{n} \mathcal{R}_{mn}\,.
\label{on-shell1}\end{equation}
The last equality follows from \eqref{5d_10d_null} and \eqref{5d_10d_eom}. A detailed discussion for the on-shell constraint in $\mathbf{5}$-representation is in Appendix \ref{section:B.2}.

Substituting the KS ansatz into \eqref{on-shell1}, we get an on-shell constraint, which is analogous to
\begin{equation}
  \partial_{p[m}\Delta_{m']}{}^{p} \partial^{q[m}\Delta_{q}{}^{m']}{} + \Delta_{m}{}^{n} \mathcal{R}^{(0)}{}_{n}{}^{m} = 0\,,
\label{}\end{equation}
where $\mathcal{R}^{(0)}{}_{m}{}^{n}$, defined in below \eqref{zeroth_eom}, is the zeroth order in $\kappa$-expansion of the equations of motion. We assume the 7-dimensional metric $G_{ij}$ satisfies $\mathcal{R}^{(0)} = 0$. The remaining nontrivial part is 
\begin{equation}
  \partial_{p[m}\Delta_{m']}{}^{p} \partial^{q[m}\Delta_{q}{}^{m']}{} = 0\,.
\label{10d_on-shell_1}\end{equation}
If we introduce $V_{mm'} = \partial_{p[m}\Delta_{m']}{}^{p}$, \eqref{10d_on-shell_1} implies $V_{mm'}$ is a null vector. 
Since $\Delta_{m}{}^{n} \sim \ell_{m}\ell^{n}$, where $\ell_{m}$ is a null vector, one can check that $\ell^{m} V_{mm'} = 0$ and 
\begin{equation}
  \Delta_{m}{}^{n} V_{np} = 0\,.
\label{}\end{equation}
Since we have constrained the totally isotropic subspace $N$ in \eqref{null_vector} to the reduced null space $\hat{N}$, the null vector $V_{mm'} \in \hat{N}$ is written in terms of two $\mathbf{5}$-vectors
\begin{equation}
  V_{mm'} = \ell_{[m} v_{m']}\,, 
\label{10_on-shell_constraint}\end{equation}
where $v$ is a vector that is orthogonal to $\ell$, $\ell_{m} m_{0}{}^{mn} v_{n} = 0$. We refer this condition as the \textit{on-shell constraint}, because it is derived from the equations of motion.

Next we expand the equations of motion $\mathcal{R}_{mm',nn'}=0$ with respect to the expansion parameter $\kappa$. One can show that $\mathcal{R}_{mm',nn'}$ is up to second order in $\kappa$ and $\mathcal{R}^{(2)}$ is the highest order, if we use the null condition of $V_{mm'}$ \eqref{10d_on-shell_1}, 
\begin{equation}
  \mathcal{R}_{mm',nn'} = \sum_{p=0}^{2}{}\kappa^{p}\mathcal{R}^{(p)}{}_{mm',nn'}\,.
\label{}\end{equation}
where
\begin{equation}
\begin{aligned}
  \mathcal{R}^{(0)}_{mm',nn'} &= -\frac{2}{3} \mathcal{M}_{0m m', p [n} \mathcal{R}^{(0)}_{n']}{}^{p}\,,
  \\
  \mathcal{R}^{(1)}_{mm',nn'} &=  \frac{1}{6 \sqrt{ |G| } } \mathcal{M}_{0 m m', p [n|} \Big[
 \partial^{qq'} \big( \sqrt{ |G| } \partial_{qq' } ( \varphi \Delta_{|n']}{}^{p}) \big) + 2 \partial_{p' r} \big(  \sqrt{ |G| } \partial^{p p'} (\varphi \Delta_{|n']}{}^{r}) \big)
 \\
  &\qquad\qquad\qquad\qquad\qquad - 2 \partial_{n']  r} \big( \sqrt{ |G| } \partial^{pp'} ( \varphi \Delta_{p'}{}^{r}) \big)
  + 4 \partial^{r [p} \big(  \sqrt{ |G| } \partial_{|n'] p'} ( \varphi \Delta_{r}{}^{p']}) \big) \Big]
  \\
  &\quad+ \frac{1}{3\sqrt{|G|}}\mathcal{M}_{0mm',p[n} \partial_{i}\Big(\sqrt{|G|} G^{ij} \partial_{j} \Delta_{n']}{}^{p}\Big) +\cdots\,,
  \\
  \mathcal{R}^{(2)}_{mm',nn'} &= \frac{1}{\sqrt{|G|}} \mathcal{M}_{0mm',pn}\Big[ \partial^{rs} \big(\sqrt{|G|} \Delta_{[r}{}^{p} V_{n's]}\big) +\frac{2}{3} \sqrt{|G|} \big(\partial^{pq}\Delta_{[q}{}^{r}\big)V_{n']r}\Big]
  \\
  &\quad  + R^{(1)}_{mm'np} \Delta_{n'}{}^{p} -(n\leftrightarrow n')\,,
\end{aligned}\label{}
\end{equation}
where $\mathcal{R}^{(0)}{}_{m}{}^{n}$ in $\mathcal{R}^{(0)}_{mm',nn'}$ is defined by
\begin{equation}
  \mathcal{R}^{(0)}{}_{m}{}^{n} = - \partial_{mp}\partial^{np} \ln \sqrt{|G|} +\frac{1}{4}\partial_{mp} G^{ij} \partial^{np} G_{ij}\,.
\label{zeroth_eom}\end{equation}
Finally, we further assume that the external $7$-dimensional metric $G_{ij}$ satisfies $\mathcal{R}^{(0)}_{mm',nn'}=0$ or $\mathcal{R}^{(0)}_{m}{}^{n}=0$. This is a consistent condition for $G_{ij}$. Note that $G_{ij}$ is assumed to be a nondynamical field, but it must still obey the consistency condition. The ellipsis in $\mathcal{R}^{(1)}_{mm',nn'}$ denotes terms that are proportional to $\mathcal{R}^{(0)}_{mm',nn'}$ and do not contribute to the equations of motion. We can show that $\mathcal{R}^{(2)}_{mm',nn'}$ vanishes when we apply the on-shell constraint for $V_{mm'}$ in \eqref{10_on-shell_constraint}. Then the remaining equation of motion is simply
\begin{equation}
  \mathcal{R}_{mm',nn'} = \mathcal{R}^{(1)}_{mm',nn'} =0\,.
\label{}\end{equation}
%

We now decompose the $\mathbf{10}$ indices in $\mathcal{R}_{mm',nn'}$ using the M-theory section condition,
\begin{equation}
  \partial_{m m'} = \begin{pmatrix} \partial_{\mu5} \\ \partial_{\mu\mu'} \end{pmatrix} \equiv \begin{pmatrix} \partial_{\mu} \\ 0 \end{pmatrix}\,, \qquad \partial^{m m'} = \begin{pmatrix} \partial^{\mu5} \\ \partial^{\mu\mu'} \end{pmatrix} \equiv \begin{pmatrix} \eta^{\mu\nu} \partial_{\nu}=\partial^{\mu} \\ 0 \end{pmatrix}\,.
\label{}\end{equation}
Each component reads
\begin{equation}
\begin{aligned}
  \mathcal{R}_{\mu5,\nu5} &= - \frac{1}{6 \sqrt{|G|}} \partial_{\sigma} \Big( \sqrt{|G|}  \big( \partial^{\sigma}  \Delta_{\mu \nu} - 2  \partial_{(\mu} \Delta_{\nu)}{}^{\sigma}\big) \Big) - \frac{1}{3 \sqrt{|G|} } \partial_{(\mu} \big( \sqrt{|G|} \partial_{\nu )}  \Delta_{55} \big) 
  \\
  &\quad - \frac{1}{6 \sqrt{|G|}} \partial_{i} \big( \sqrt{|G|} G^{i j} \partial_{j} \Delta_{\mu \nu} \big) 
  \\
  \mathcal{R}_{\mu5,\nu\nu'} &= -\frac{1}{3\sqrt{|G|}} \eta_{\mu [\nu} \partial_{\nu']} \big(\sqrt{|G|} \partial^{\rho} \Delta_{\rho}{}^{5} \big) - \frac{1}{3 \sqrt{|G|} } \eta_{\mu [\nu| } \partial_{i} \big(\sqrt{|G|} G^{i j} \partial_{j}  \Delta_{|\nu']}{}^{5}\big) 
\end{aligned}\label{eom_component}
\end{equation}
Since $\mathcal{M}^{\mu5,\nu5} = |g|^{-\frac{1}{5}}g^{\mu\nu}$ and $\mathcal{M}^{\mu5,\nu\nu'} =\frac{1}{2} |g|^{-\frac{1}{5}}g^{\mu\rho}C_{\rho\sigma\sigma'}\epsilon^{\nu\nu'\sigma\sigma'}$, we can identify $\mathcal{R}_{\mu5,\nu5}$ and $\mathcal{R}_{\mu5,\nu\nu'}$ as the equations of motion for $g$ and $C$ respectively. We omit $\mathcal{R}_{\mu\mu',\nu\nu'}$, because it can be written by a combination of $\mathcal{R}_{\mu5,\nu5}$ and $\mathcal{R}_{\mu5,\nu\nu'}$.

\section{Kerr-Schild Double copy for M-theory}
The double copy structure or colour-kinematic duality \cite{Bern:2008qj,Bern:2010ue,Bern:2010yg} for tree level scattering amplitudes is equivalent to the KLT relation \cite{Kawai:1985xq} in closed string scattering amplitude. Thus there is a clear picture of the double copy structure in the framework of the closed string theory. It stems from the left and right mode decomposition in the closed string mode expansion, and each left and right mode corresponds to a gauge field in infinite tension limit or field theory limit, $\alpha'\to 0$. 

M-theory is the strong coupling limit of the type IIA string theory, and there is no string like object in 11-dimensional spacetime. Its degrees of freedom are given by the M-branes, the membranes and fivebranes. Quantisation of these M-branes is significantly different from quantising strings, and there is certainly no equivalent of left-right mode decomposition which is a peculiarity of the string. Therefore it is not obvious whether such a the double copy structure exists in M-theory and its low energy effective description, 11-dimensional supergravity.

In this section, we consider the Kerr-Schild classical double copy for M-theory (in contrast to the double copy for scattering amplitudes). The Kerr-Schild double copy yields an exact double copy relation for the solutions of the equations of motion that are consistent with the universal KS ansatz. In waht follows, the field equations of the generalised metric under the KS ansatz will be reduced to linear equations for a Maxwell field and two form. We can use this result to establish an exact double copy map. The advantage of classical double copy relation is that it is much simpler to establishing a map as compared to scattering amplitudes. Once this classical map has been established it begs the question as to whether this classical double copy relation might be extended for scattering amplitudes and that eleven dimensional supergravity will obey a full double copy relation. Finally, we further discuss the Kerr-Schild double copy for type-IIB supergravity by choosing the so called ``IIB section'' in exceptional field theory. This then allows us to construct a double copy relation for the fields of the IIB supergravity including both the NS and RR two forms. 

\subsection{M-theory section}
Before considering the KS double copy relation for $SL(5)$ ExFT, we want to rewrite the equations of motion \eqref{eom_component} in terms of the components of $K_{mm'}$ in \eqref{comp_null}, $l_{\mu}$ and $k_{\mu\nu}$, by using the relation in \eqref{Delta}. Substituting the parametrization of $K_{mm'}$, $\mathcal{R}_{\mu5,\nu5}$ reduces to
\begin{equation}
\begin{aligned}
  \mathcal{R}_{\mu5,\nu5} &= - \frac{1}{18 \sqrt{|G|}} \partial_{\sigma} \Big( \sqrt{|G|}  \big( \partial^{\sigma}  (\varphi l_{\mu} l_{\nu} - \varphi \tilde{k}_{\mu \rho} \tilde{k}_{\nu}{}^{\rho}) - 2  \partial_{(\mu} (\varphi l_{\nu)}l^{\sigma} -\varphi \tilde{k}_{\nu)\rho} \tilde{k}^{\sigma\rho})\big) \Big) 
  \\
  &  - \frac{1}{9 \sqrt{|G|} } \partial_{(\mu} \big( \sqrt{|G|} \partial_{\nu )}  (\varphi l_{\rho} l^{\rho}) \big) - \frac{1}{18 \sqrt{|G|}} \partial_{i} \big( \sqrt{|G|} G^{i j} \partial_{j} (\varphi l_{\mu}l_{\nu} - \varphi\tilde{k}_{\mu\rho} \tilde{k}_{\nu}{}^{\rho}) \big)=0	\,.
\end{aligned}\label{eom1}
\end{equation}
Similarly, we also recast $\mathcal{R}_{\mu5,\nu\nu'}$ in terms of $l_{\mu}$ and $k_{\mu\nu}$. After contracting $\eta^{\mu\nu}$ with $\mathcal{R}_{\mu5,\nu\nu'}$ and using the following identity,
\begin{equation}
  \epsilon^{\mu\rho_{1}\rho_{2}\rho_{3}} \partial_{\mu}\big(\epsilon_{\nu_{1}\nu_{2}\nu_{3}\nu_{4}} X^{\nu_{1}\nu_{2}\nu_{3}\nu_{4}}\big) = 4!\, \partial_{\mu}\big(X^{[\mu\rho_{1}\rho_{2}\rho_{3}]}\big)\,,
\label{}\end{equation}
and we have
\begin{equation}
\begin{aligned}
 \mathcal{R}_{\mu5,\nu\nu'} \longrightarrow \frac{4}{\sqrt{|G|}} \partial^{\lambda} \Big( \sqrt{|G|} \partial_{[\lambda} ( \varphi l_{\mu} k_{\nu\rho]}) \Big) + \frac{1}{\sqrt{|G|} } \partial_{i} \Big( \sqrt{|G|} G^{i j} \partial_{j} (\varphi l_{[\mu} k_{\nu \rho]} ) \Big) =0\,.
\end{aligned}\label{eom2}
\end{equation}

Let us now suppose that the $4$-dimensional spacetime admits at least one Killing vector $\xi$. Take a coordinate $x^{\mu} = \{y,x^{\alpha}\}$ that the Killing vector becomes constant, then it is represented as $\xi^{\mu} = \frac{\partial}{\partial y} = \delta^{\mu}_{y}$. In this coordinate system, there is no $y$-coordinate dependence for all the fields, $\partial_{y} = 0$. Since our main interest is the 4-dimensional subspace, we further assume that the external metric is trivial such as $\sqrt{|G|} = 1$, and we drop the dependence on the 7-dimensional external directions, $\partial_{i} = 0$, by considering smeared solutions only. After contracting the Killing vector with the above equations \eqref{eom1} and \eqref{eom2}, we have
\begin{equation}
\begin{aligned}
  &\partial^{\sigma} \partial_{\sigma} \Big(\varphi l_{\mu}(\xi\cdot l) - \varphi \tilde{k}_{\mu\rho} (\xi\cdot \tilde{k})^{\rho} \Big)-\partial^{\sigma} \partial_{\mu} \Big( \varphi (\xi\cdot l)l_{\sigma}  -  \varphi \tilde{k}_{\sigma\rho} (\xi\cdot \tilde{k})^{\rho} \Big) =0\,,
\\
  &\partial^{\lambda} \partial_{[\lambda} \big( \varphi (\xi \cdot l)k_{\nu\rho]} +\varphi l_{\nu}(\xi\cdot k)_{\rho]} \big)= 0\,,
\end{aligned}\label{}
\end{equation}
where $(\xi\cdot \tilde{k})_{\nu} =\xi^{\mu} \tilde{k}_{\mu\nu}$. 

We may separate the first equation into two parts with respect to $\varphi l_{\mu}$ and $\varphi \tilde{k}_{\mu\nu}$. We will take these to vanish seperately. In fact, to be consistent with the single copy of the type IIA supergravity (discussed below) this will have to be the case. Here we will focus on $\varphi l_{\mu}$ part
\begin{equation}
  \partial^{\sigma} \partial_{\sigma} \Big(\varphi l_{\mu}(\xi\cdot l) \Big)-\partial^{\sigma} \partial_{\mu} \Big( \varphi (\xi\cdot l)l_{\sigma}\Big) =0\,.
\label{}\end{equation}
We impose conditions on $l_{\mu}$ and $k_{\mu\nu}$ as
\begin{equation}
  \xi \cdot l = \mbox{constant}\,,\qquad \xi^{\mu} k_{\mu\nu} = 0\,,
\label{}\end{equation}
and we have
\begin{equation}
\begin{aligned}
  &\partial^{\sigma} \partial_{\sigma} \big(\varphi l_{\mu}\big)-\partial^{\sigma} \partial_{\mu} \big( \varphi l_{\sigma}\big) =0\,,
  \\
  &\partial^{\rho} \partial_{[\rho} \big(\varphi k_{\mu\nu]}\big) = 0\,.
\end{aligned}\label{}
\end{equation}
\begin{equation}
\begin{aligned}
  A_{\mu} &=  \varphi l_{\mu}\,,   \qquad  B_{\mu\nu} &= \varphi k_{\mu\nu} \,.
\end{aligned}\label{}
\end{equation}
This implies that $\mathcal{M}_{mm',nn'}$ may be written in terms of the 1-form and 2-form fields. 

The single copy of the NS-NS sector of  the IIA theory string has already been described in \cite{Lee:2018gxc}. It is given by two Maxwell fields $\mathcal{A}_{\mu}$ and $\bar{\mathcal{A}}_{\mu}$. A natural check on the formalism here is to dimensionally reduce this M-theory formalism on a circle and relate it to the single copy for the string. If we take $x^2$ to be the ``M-theory'' circle on which we are reducing then the IIA single copy fields may be written in terms of the M-theory single copy fields as follows:
\begin{equation}
  \mathcal{A}_{\mu} = \frac{1}{2}(A_{\mu} + B_{\mu2}) \,,\qquad \bar{\mathcal{A}}_{\mu} =\frac{1}{2}( A_{\mu} - B_{\mu2})\, .
\label{IIA_SC}\end{equation}
%

\subsection{Type IIB section}

ExFT admits two independent choices of {\it{section}}. The so called M-theory section, which is the most immediate choice we describe above has a spacetime described by the coordinates $x^\mu$, with $\mu=1..4$ and all fields are independent of the other coordinates.
The IIB section choice on the other hand has spacetime described by the coordinates: $x_{ab}$ with $a=1..3$ and all fields are then taken to be independent of the other coordinates. 

This is then a three dimensional section choice that ultimately related to the IIB supergravity. The seven dimensional Riemannian space goes along for the ride, just as before, providing a total of ten dimensions as needed. In what follows we will follow \cite{Blair:2013gqa} where the IIB section choice was first constructed for the $SL(5)$ theory. It is useful to describe IIB coordinates as follows $\tilde{x}_a= \frac{1}{2} \epsilon_{abc} x^{bc}$. In terms of the $\bf{5}$ representation of $SL(5)$ one decomposes to $SL(3)\oplus SL(2)$ or in terms of indices $m \rightarrow (a,i)$ with $a$ an $SL(3)$ index and $i$ an SL(2) index (In what follows the indices $a,b,c,d,e,f$  will be $SL(3)$ indices and run $1..3$ and $i,j,k,l$ will be $SL(2)$ indices and run $1, 2$). The IIB choice of section may then be described by demanding that all fields in the theory obey:
\begin{equation}
\frac{\partial}{\partial x^{ai}}=0 \, ,\qquad \frac{\partial}{\partial x^{ij}}=0 \, .
\end{equation}

The generalised ``little metric'' adapted for this choice of section is then:
\begin{equation}
  m_{mn} = |g|^{-\frac{1}{10}} \begin{pmatrix} {|g|}^{-\frac{1}{2}} (g_{a b} + \frac{1}{4} |g|^{\frac{1}{2}} H_{ij} \epsilon_{acd} \epsilon_{bef} C^{cdi} C^{efj}) & {|g|}^{{\frac{1}{2}}} H_{jk} \epsilon_{bef} C^{efk} \\ {|g|}^{{\frac{1}{2}}} H_{il} \epsilon_{acd} C^{cdl} & ~~-{|g|}^{\frac{1}{2}}H_{ij} \end{pmatrix} \, ,
\label{}\end{equation}
with the $SL(2)$ metric $H_{ij}$ given by:
\begin{equation}
H_{ij}= \frac{1}{\tau_2} \begin{pmatrix} 1 & \tau_1 \\ \tau_1 & \tau_1^2 + \tau_2^2 \end{pmatrix} \, . \label{slmetric}
\end{equation}

Of course one can introduce the $SL(2)$ doublet of one form fields $v_{a}^i$ as follows:
\begin{equation}
v_a^i= \frac{1}{2} \epsilon_{abc} C^{ibc} \, 
\end{equation}
and then use this pair of one forms.

We will carry out the Kerr-Schild Ansatz for this IIB section choice. The ansatz for $m_{mn}$ will be the same as before, namely:
\begin{equation}
  m_{mn} = m_{0mn} + \kappa \Delta_{mn}\,, \qquad \qquad \Delta_{mn} = \varphi \ell_{m} \ell_{n}\,,
\label{}\end{equation}
and again $\ell_{m}$ is a null vector in $\mathbf{5}$-representation. 

To carry out the IIB choice we now decompose $\ell_{m}$ as follows:
\begin{equation}
\ell_m= (\ell_a,\ell_i) \, .
\end{equation}
We will take our $m_{0mn}$ to be flat Minkowski space with signature $(+,-,-)$ with our sign choices exactly following \cite{Blair:2013gqa} where the Lorenzian IIB theory is described using the $\mathbf{5}$-representation. (One can also take a more complete ansatz as with the M-theory case and introduce another null vector field $\ell'_m$ orthogonal to $\ell_m$.) The null condition on $\ell_m$ becomes:
\begin{equation}
\ell_a \eta^{ab} \ell_b + \ell_i \ell^i =0 \, .  \label{nulll}
\end{equation}
If the $SL(2)$ doublet vanishes then we have a simple null condition on $\ell_a$ and one will recover the usual Kerr-Schild Ansatz for gravity alone. This is analogous to setting $l_5=0$ in the M-theory section. Solving (\ref{nulll}) for non zero $\ell_i$ is now more involved than the previous case since we have a whole $SL(2)$ orbit of choices  for $\ell_i$.

Next we relate the ansatz to supergravity fields of IIB supergravity (on the 3d space) $g_{ab}, B_{ab}, C_{ab}, \phi, C_0$. First one writes the NS two form, $B_{ab}$ and RR two form, $C_{ab}$ as a doublet $C_{ab}^i$ where i=1 denotes the NS two form and i=2 denotes the RR two form. It is immediate that the components of the $SL(2)$ metric may be identified with the axion-dilaton as follows:
\begin{equation}
C_0 + i e^{-\phi}  = \tau_1 +i \tau_2  \, .
\end{equation}
Then we find:

\begin{equation}
\begin{aligned}
  |g| &= (1 -\kappa\Delta_{i}{}^{i})^{-\frac{5}{4}} 
  \\
  g_{a b} &= (1 - \kappa\Delta_{i}{}^{i})^{\frac{3}{4}} \Big(\eta_{ab}- \kappa\Delta_{ab}-{\kappa^{2}}{(1-\kappa \Delta_{j}{}^{j})^{\frac{1}{2}}} \Delta_{a i} \Delta^{i}{}_{b} \Big) 
  \\
  g^{ab} &= (1 - \kappa \Delta_{i}{}^{i})^{-\frac{3}{4}}  \big(\eta^{ab} + \kappa\Delta^{ab} \big)  
  \\
  C^{ab j} &= {(1 -\kappa\Delta_{i}{}^{i})^{\frac{1}{2}}} \epsilon^{abc} \Delta^{j}{}_{c} \,.
\end{aligned}\label{KS_componentsiib}
\end{equation}

One can now relate the IIB supergravity fields to the components of the null one form $\ell_a,\ell_i$ in $\Delta_{mn}$  using this dictionary.

In this paper we will not repeat the full calculation using the {\bf{10}} representation which becomes somewhat tedious but simply note the decomposition of the {\bf{10}} vector as follows: $K_M=K_{mn} \rightarrow K_{ia}, K_{ij}, K_{ab}$. The components $K_{ia}$ then provide the single copy one forms required to make the $SL(2)$ doublet of two-forms and the components $K_{ab}$ can be used to make a single copy $SL(2)$ singlet one-form. We may also follow the calculation in appendix B, which gives the equations of motion using the {\bf{5}} representation and thus have the equations of motion but now we use the  IIB decomposition for the null vector $\ell$. To get the single copy equations we make similar simplifications as above for the M-theory section and take $\det |G| = 1$ and thus consider smeared solutions along the external directions. Finally, when the dust settles one has three sets of Maxwell equations for the IIB single copy. From \eqref{Delta}, $\Delta$ is given by $K_{mm'}$ as follows:
\begin{equation}
  \Delta_{ab} = \frac{\varphi}{3} \big(K_{ac} K_{b}{}^{c} + K_{ai} K_{b}{}^{i}\big) \,, \qquad \Delta_{ai} = \frac{\varphi}{3} \big(K_{aj} K_{i}{}^{j} + K_{ab} K_{i}{}^{b}\big)\,.
\label{}\end{equation}
Then there are two Maxwell fields, $SL(2)$ singlet and doublet
\begin{equation}
  A_{a} = \varphi \epsilon_{abc} K^{bc} \,, \qquad A^{i}_{a} = \varphi K_{ai} \, .  
\label{IIBsinglecopy}\end{equation}
This is what we would expect. For the case where there are no two-form fields present, the single copy for gravity is described by the $SL(2)$ singlet Maxwell field. The two-forms of IIB may be constructed from the $SL(2)$ doublet Maxwell one-forms in the single copy just as in \cite{Lee:2018gxc}.

\section{Charged Brane solutions}
We will now construct the known electrically charged brane solutions \cite{Berman:2007bv} in terms of the universal Kerr-Schild Ansatz and interpret these solutions from the point of view of the single copy fields. From the exceptional field theory perspective these solutions have already been given a wave-like construction \cite{Berman:2014hna} and so in fact the Kerr-Schild type construction is very natural.

\subsection{M-theory section and the Membrane}
Let's assume the worldvolume directions of the M2-brane are $t, x^{1},x^{2}$ and denote them as $x^{\alpha} = \{t,x^{1},x^{2}\}$ and choose the M-theory circle direction as $x^{2}$. The transverse directions are $\vec{x}_{8} = \{x^{3},z^{i}\}$, where $i,j,\cdots$ are 7-dimensional extra directions. The M2-brane geometry is given by
\begin{equation}
\begin{aligned}
  &\mathrm{d}s^{2}_{11} = H^{-\frac{2}{3}} \eta_{\alpha\beta} \mathrm{d}x^{\alpha} \mathrm{d}x^{\beta} + H^{\frac{1}{3}} \mathrm{d}x^{3} \mathrm{d}x^{3}+ H^{\frac{1}{3}} \delta_{ij} \mathrm{d}z^{i} \mathrm{d}z^{j}\,,
  \\
  &C_{t12} = -\big(1-H^{-1}\big)\,,\qquad H= 1+\frac{h}{|\vec{x}_{8}|^{6}}\,.
\end{aligned}\label{}
\end{equation}
To embed the 11-dimensional supergravity into the $SL(5)$ ExFT, we need to use the following Kaluza-Klein ansatz for the 11-dimensional metric $\hat{g}$ \cite{Berman:2019izh}
\begin{equation}
  \hat{g}_{\hat{\mu} \hat{\nu}}=\left(\begin{array}{cc}|g|^{-\frac{1}{5}} G_{ij} + A_{i}^{\rho} A_{j}^{\sigma} g_{\rho\sigma} & A_{i}^{\rho} g_{\rho\nu} \\  g_{\mu k} A_{j}^{k} & g_{\mu\nu}\end{array}\right)\,,
\label{bigmet}\end{equation}

For the M2-brane example, the external gauge field $A = 0$ and the seven dimensional external metric $G_{ij}$ is trivial
\begin{equation}
  G_{ij} = \delta_{ij}\,.
\label{}\end{equation}
The internal metric and $C$-field are
\begin{equation}
  g_{\mu\nu} = H^{-\frac{2}{3}} \big( \eta_{\alpha\beta} \mathrm{d}x^{\alpha} \mathrm{d}x^{\beta} + H\mathrm{d}x^{3}\mathrm{d}x^{3}\big)\,, \qquad C_{012} = H^{-1} -1  \, .
\label{}\end{equation}
The most efficient way to encode this solution is using the  $\mathbf{5}$-representation, where the generalized metric is 
\begin{equation}
  m_{mn} = \begin{pmatrix} \eta_{\alpha\beta} & 0 & 0 \\ 0 & H & 1-H \\ 0 & 1-H & H-2\end{pmatrix}
\label{}\end{equation}
and it can be written in terms of the KS-ansatz with a single null vector $\ell$
\begin{equation}
  m_{mn} = m_{0mn} +\kappa \varphi \ell_{m} \ell_{n} \,,
\label{}\end{equation}
where
\begin{equation}
  m_{0mn} = \begin{pmatrix} \eta_{\alpha\beta} & 0 & 0 \\ 0 & 1 & 0 \\ 0 & 0 & -1\end{pmatrix}\,,\qquad \kappa  \varphi = H-1\,, \qquad \ell_{m} = \begin{pmatrix} \mathbf{0}_{3} \\ 1 \\ 1 \end{pmatrix}\,.
\label{5solution}\end{equation}
From this, using the map provided between the $\mathbf{5}$ representation and the supergravity fields one may recover the the full membrane solution.

When we convert the $\mathbf{5}$ Kerr-Schild ansatz to $\mathbf{10}$ representation we need the auxiliary field $j$ which is defined by
\begin{equation}
  \qquad j_{m} = \begin{pmatrix} 2\sqrt{3} \\ 0 \\ \mathbf{0}_{3}\end{pmatrix}\,,
\label{}\end{equation}
which satisfy the requirements that $\ell \cdot j = 0$ and $j\cdot j = -12$. Then we can read off the $\mathbf{10}$ null vector components
\begin{equation}
\begin{aligned}
    K_{05} &= \ell_{[0} j_{5]} = l_{0} = \sqrt{3} \,, \qquad  K_{03} = \ell_{[0} j_{3]} = \tilde{k}_{03} = k_{12} = -\sqrt{3}\,, 
\end{aligned}\label{}
\end{equation}
or
\begin{equation}
  l_{\mu} = \big(\sqrt{3},0,0,0\big)\,, \qquad k_{\mu\nu} = \begin{pmatrix} 0 & 0 & 0 & 0\\0 & 0 & -\sqrt{3} & 0\\ 0 & \sqrt{3} & 0 & 0 \\ 0 & 0 & 0 & 0\end{pmatrix}
\label{}\end{equation}
with
\begin{equation}
\kappa  \varphi = H-1 \, ,
\end{equation}
obeys the Poisson equation in codimension one. Note that $\xi^{\mu}k_{\mu\nu} = k_{0\nu} =  0$ for the timelike Killing vector $\xi = \partial_{0}$.  The single copy of this solution is then given by: 
\begin{equation}
A_0 = \varphi  \,, \qquad B_{12}= \varphi\,,
\end{equation}
where we have rescaled to satisfy $\xi\cdot l = 1$.

Thus the single copy of the membrane is described by a two dimensional plane of electric charges. That is the Maxwell field is given by solving the Harmonic equation for an electric source smeared over two spatial dimensions just like the classical electrostatic problem for a plane of charge. The two form solution is like that of a magnetic string that is smeared over one additional dimension to give a plane of string charge.
Both the Maxwell field and the two form field are given in terms of the same single harmonic function $\varphi$. This is due to the BPS nature of these solutions. If there were no $C$ field and then the membrane was purely a gravitational object then its solution would be described by the usual gravitational single copy relation by the Maxwell field. The two form encodes the $C$ field contribution. The fact being BPS sets the charge equal to the mass (or tension) of the solution means that what would be independent Harmonic functions for the Maxwell and two form fields are fixed to be equal. 
We thus conjecture that the non-BPS solutions are described by the same Ansatz but with independent Harmonic functions for $A$ and $B$ fields. We leave this for future research.

Finally we make a comment the type IIA reduction. The above result is consistent with the single copy of the fundamental string solution in type IIA theory. According to the relation in \eqref{IIA_SC}, the corresponding type IIA single copies are
\begin{equation}
  \mathcal{A} = H \begin{pmatrix} 1\\1\\\mathbf{0}\end{pmatrix}\,, \qquad \bar{\mathcal{A}} = H\begin{pmatrix} 1\\-1\\\mathbf{0}\end{pmatrix}\,.
\label{}\end{equation}
%

\subsection{Type IIB section: the fundamental and D strings}
We now will take the alternative choice of section to provide a description of IIB supergravity and examine the  related IIB solutions which describe the fundamental and D strings.
We will work directly with the $\mathbf{5}$ and write the solution for the IIB section as:
\begin{equation}
\ell_0=0, \,  \ell_1=0,  \, \ell_2=1 \, , \qquad \ell_i=(1,0) \, \qquad \kappa \varphi=H-1
\end{equation}
with $H$ solving the Harmonic equation in codimension one.
Thus, using (\ref{IIBsinglecopy}) the single copy is given by:
\begin{equation}
A_0=\varphi    \, \qquad A^1_2= \varphi   \, \qquad A_a^2=0 \, .
\end{equation}
This is a Maxwell field sourced by a point electric charge smeared along a line in two spatial dimensions with another $SL(2)$ Maxwell field sourcing a magnetic charge again smeared over a line.
Following the dictionary described above in (\ref{KS_componentsiib}) we can then write the solution in terms of supergravity fields as follows, where we use the coordinates $x^\alpha, (\alpha=0,1), x^3, z^i, (i=1..7)$:
\begin{equation}
\begin{aligned}
  ds^2_{IIB} &= H^{-\frac{3}{4}} \big( \eta_{\alpha\beta} \mathrm{d}x^{\alpha} \mathrm{d}x^{\beta} + H\mathrm{d}x^{3}\mathrm{d}x^{3}\big)+ H^{\frac{1}{4}} \delta_{ij}dz^idz^j \,,\\  B_{01} &= H^{-1} -1  \, ,  \qquad C_{(2)}=0 \,.
\end{aligned}\label{}
\end{equation}
We may immediately identify this as the fundamental string written in Einstein frame. It is particularly satisfying to see how the different powers of the Harmonic function, H emerge from the nonlinear expressions in (\ref{KS_componentsiib}) and (\ref{bigmet}).
If we now make the alternative choice for the $SL(2)$ doublet $\ell_o=(0,1)$ then the metric is the same but the RR field is no longer zero and we obtain:
\begin{equation}
C_{01} = H^{-1} -1  \, ,  \qquad B_{(2)}=0 \,,
\end{equation}
which we may identify as the D1 brane in Einstein frame.
Of course one can take a linear combination for our $SL(2)$ doublet of these solutions such that $l_a=(p,q)$ and one obtains the solution of the $(p,q)$ string. The expressions for the axion-dilaton are given as expected by the $SL(2)$ metric.

\section{Conclusion}
We have constructed the classical double copy for M-theory by using the ExFT formalism with a Kerr-Schild Ansatz for the generalized metric. Although we focussed on the $SL(5)$ ExFT we have developed a universal Kerr-Schild ansatz (including the known examples in General Relativity and Double Field Theory). This ansatz when applied to the $SL(5)$ ExFT produces linear equations of motion! 

Making manifest this hidden linearity is an essential part of the classical double copy story. The double copy for M-theory that we have written down (specifically for the SL(5) ExFT) gives the exact relation between the solutions of 11-dimensional supergravity and free 1-form and 2-form field theories in four dimensions. As with the usual KS double copy, it only covers a limited region of the full solution space where there is sufficient symmetry. In fact, seeking the full double copy map is a highly nontrivial task due to the dependence on gauge choices and field redefinitions. For gravity in four dimension, there is also a perturbative approach for the classical double copy \cite{Luna:2016hge,Goldberger:2016iau,Goldberger:2017frp,Goldberger:2017vcg,Goldberger:2017ogt,Carrillo-Gonzalez:2018pjk,Shen:2018ebu,Plefka:2018dpa}. The strength of this perturbative method is its generality, it allows us to make statements about relations for generic perturbative solutions. It would be intriguing to apply the perturbative method to the ExFT to understand the full double copy relation for M-theory. Further, given the specification of single copy fields we have made it is tempting to consider a possible double copy relation for the scattering amplitudes in eleven dimensional supergravity.

We have presented the membrane solution in M-theory and F1/D1 strings in type IIB as examples using this Kerr-Schild ansatz. One limitation is that due to using the $SL(5)$ ExFT we required the solutions to have additional ``smearing" so that there were additional isometries in the non-generalised space. This led to a slight pathology in that the Harmonic functions in play were only of codimension one or two. Going to exceptional field theories of higher rank will fix this and the associated harmonic function will then be for higher codimension and be better behaved asymptotically.
One may also consider the Killing spinor equation within the framework of KS formalism in the supersymmetric version of $SL(5)$ ExFT, which may lead to linear first order partial differential equations as in  \cite{Lee:2018gxc,Lescano:2020nve}.

In addition, it should be relatively straightforward to use the universal Kerr-Schild ansatz and consider the corresponding classical double copy for all the Exceptional Field Theories associated to the various exceptional groups, $SO(5,5)$, $E_{6(6)}$, $E_{7(7)}$ and $E_{8(8)}$. Each one of these will have a set of linear single copy fields providing a specific map between the supergravity fields and the single copy field in various dimensions. The hierarchy presented in this paper where the three form leads to a two form single copy field (and the previously discovered relation that the two form potential gives rise to a one form in the single copy) suggests an interesting general structure may emerge that relates p-forms in the double copy to (p-1)-forms  in the single copy. This possible relationship to ``higher structures'' motivates us further to look into the universal Kerr-Schild Ansatz for all the Exceptional Field Theories.

As a further extension, it is known that a similar classical double copy structure exists for the so called multi Kerr Schild Anstaz which leads to the obvious extension of this work to a universal multi Kerr Schild Ansatz. Such Ansatze will be necessary to capture more diverse spacetimes such as superpositions of branes. As mentioned previously, the BPS nature of the solutions in this paper means that we have sufficient symmetry in the solutions such that all the data is captured by one Harmonic function. The non-BPS brane  solutions require two Harmonic functions, which can be accomodated within this formalism with a simple extension where the B-field and A-field have different harmonic solutions. All of these possibilities show that this paper is just the first step along the road to describing large classes of solutions in eleven dimensional supergravity as generalised Kerr-Schild solutions with a single copy description.

\section*{Acknowledgments}
DSB is supported by the UK Science and Technology Facilities Council (STFC) with consolidated grant ST/L000415/1, {\it{String Theory, Gauge Theory and Duality}}. KL is supported by an appointment to the JRG Program at the APCTP through the Science and Technology Promotion Fund and Lottery Fund of the Korean Government. It is also supported by the Korean Local Governments - Gyeongsangbuk-do Province and Pohang City. We thank Chris Blair, Ricardo Monteiro, Bill Spence and Chris White for discussions on this topic.


\appendix

\section{Results in \bf{5}-dimensional representation \label{sec:B}}
\subsection{KS ansatz}
Let us consider the KS ansatz for the small metric $m_{mn}$, which is related to the generalised metric in $\mathbf{10}$-representation $\mathcal{M}_{mm',nn'}$ by \eqref{little_metric}. It is parametrized in terms of 4-dimensional fields, metric $g_{\mu\nu}$, 1-form field $v_{\mu}$, hodge dual of $C_{3}$ field,
\begin{equation}
  m_{mn} = |g|^{-\frac{2}{5}} \begin{pmatrix} g_{\mu\nu} & -\sqrt{|g|} v_{\mu} \\ -\sqrt{|g|}v_{\nu} & ~~|g|\big( -1 + v^{\mu} v_{\mu}\big) \end{pmatrix} \,,
\label{para_small_gen_metric}\end{equation}
where $m,n,\cdots = 0,1,2,3\ \mathrm{and}\ 5$. The inverse and the determinant of the small metric are given by
\begin{equation}
\begin{aligned} 
  m^{mn} = |g|^{\frac{2}{5}} \begin{pmatrix} \big( g^{\mu\nu} - v^{\mu } v^{\nu} \big) &~~ -\frac{1}{\sqrt{|g|}} v^{\mu} \\ \frac{1}{\sqrt{|g|}} v^{\nu} & -\frac{1}{|g|} \end{pmatrix}\,, \qquad \det m = 1\,.
\end{aligned}
\label{}\end{equation}

In the Lorentzian signature, the structure group of the generalized frame bundle is given by $\mathit{SO}(2,3)$ \cite{Berman:2019izh,Hull:1998br}. The dimension of a null subspace is determined by the metric signature. For $SO(2,3)$ case, one can introduce a two dimensional null subspace, and this implies that there are two mutually orthogonal null vectors $\ell_{m}$ and $\ell'_{m}$ satisfying
\begin{equation}
  \ell^{m} \ell_{m} =0\,,\qquad \ell'^{m} \ell'{}_{m} =0\,,\qquad \ell^{m}\ell'{}_{m} =0 \,.
\label{}\end{equation}
However, it is not necessary to introduce these two null vectors for the KS ansatz at the same time, because the general KS ansatz in \eqref{Gen_KS_ansatz} requires only single null vector.

We now introduce a Kerr-Schild type ansatz for $m_{mn}$. Since the projection operator for $5$-representation is trivial, it is the same form with the KS ansatz for the metric tensor in GR
\begin{equation}
  m_{mn}  = m_{0 mn} + \kappa \varphi \ell_{m} \ell_{n}\,,
\label{gKS}\end{equation}
where $m_{0ab}$ is a background generalized metric which is parametrized as
\begin{equation}
\begin{aligned}
    m_{0mn} =|\tilde{g}|^{-\frac{2}{5}} \begin{pmatrix}  \tilde{g}_{\mu\nu} & -\sqrt{|\tilde{g}|} \tilde{v}_{\mu} \\ -\sqrt{|\tilde{g}|} \tilde{v}_{\nu} & ~~ |\tilde{g}| \big( -1 + \tilde{v}^{\mu} \tilde{v}_{\mu}\big) \end{pmatrix} \,.
\end{aligned}\label{KS_ansatz}
\end{equation}
It is obvious that the inverse generalized metric and its determinant take the form
\begin{equation}
\begin{aligned}
  (m^{-1})^{mn} &= (m_{0}^{-1})^{mn} - \kappa \varphi \ell^{m} \ell^{n}\,,
\end{aligned}\label{}
\end{equation}
and $\det m = \det m_{0}$. Note that the KS ansatz for the small metric is consistent with the KS ansatz for $\mathcal{M}_{mm',nn'}$. If we use \eqref{little_metric}, the generalised metric is expanded
\begin{equation}
  \mathcal{M}_{mm',nn'} = \mathcal{M}_{0mm',nn'} -2\kappa \varphi m_{0[m|n|} \ell_{m']} \ell_{n'} -2 \kappa \varphi m_{0m[n'} \ell_{|m|} \ell_{n']}\,,
\label{}\end{equation}
and it reproduces \eqref{explicit_Q} and \eqref{Delta}.

Without loss of generality the null vectors $\ell^{m}$ is parametrized as
\begin{equation}
  \ell^{m} = |\tilde{g}|^{\frac{1}{5}}\begin{pmatrix} l^{\mu} +\tilde{v}^{\mu} k \\ \frac{1}{\sqrt{|\tilde{g}}|}k\end{pmatrix}\,,\qquad 
  \ell_{m} =  |\tilde{g}|^{-\frac{1}{5}} \begin{pmatrix} l_{\mu} \\ -\sqrt{|\tilde{g}|} ( \tilde{v}\cdot l + k)\end{pmatrix} \,.
\label{ansatz_para_K}\end{equation}
One may represent the null vector in a simple way by introducing a generalized frame field $E_{a}{}^{i}$ satisfying
\begin{equation}
  m_{mn}= E_{m}{}^{a} \eta_{ab} (E^{t})^{b}{}_{n}\,,\qquad \eta_{ab} = (E^{t})_{a}{}^{m} m_{mn} E^{m}{}_{b}\,,
\label{}\end{equation}
where $a,b,\cdots = 0,1,2,3,5$ are the generalized frame indices and $\eta_{ij}$ is the metric for the local $SO(2,3)$, $\eta_{ij} = \mathrm{diag}(-1,1,1,1,-1)$. The generalized frame field is parametrized 
\begin{equation}
  E_{m}{}^{a}= |g|^{-\frac{1}{5}} \begin{pmatrix} e_{\mu}{}^{m} & 0 \\ - \sqrt{|g|} v^{\mu}e_{\mu}{}^{m} & \sqrt{|g|} \end{pmatrix} \,,
\label{}\end{equation}
where $e_{\mu}{}^{m}$ is the frame field in Riemannian geometry.

Then we can rewrite the null vectors using the background generalized frame field $E_{0}$ as
\begin{equation}
  \ell_{m} = {E}_{0 m}{}^{a} \hat{\ell}_{a}\,,\qquad   \ell'_{m} = {E}_{0 m}{}^{a} \hat{\ell}'_{a}\,,
\label{}\end{equation}
where 
\begin{equation}
  \hat{\ell}_{a} = \begin{pmatrix} (\tilde{e}^{t})_{\underline{a}}{}^{\mu} l_{\mu} \\ k\end{pmatrix}\,, \qquad \hat{\ell}'_{a} = \begin{pmatrix} (\tilde{e}^{t})_{\underline{a}}{}^{\mu} l'_{\mu} \\ k'\end{pmatrix}\,.
\label{}\end{equation}
and $\underline{a},\underline{b},\cdots $ are the frame indices for the 4-dimensional spacetime.

Substituting \eqref{ansatz_para_K} into the null conditions, we have
\begin{equation}
  l^{\mu} \tilde{g}_{\mu\nu} l^{\nu} - k^{2} = 0\,, \qquad l'^{\mu} \tilde{g}_{\mu\nu} l'^{\nu} - k'^{2} = 0\,.
\label{}\end{equation}
Assuming the $l$ and $l'$ are spacelike or null, we get the real solutions
\begin{equation}
  k = \pm \sqrt{l\cdot l}\,, \qquad k' = \pm \sqrt{l'\cdot l'}\,,
\label{}\end{equation}
Using the parametrization of the null vector, we may read off the KS ansatz for the component fields 
\begin{equation}
\begin{aligned}
  g_{\mu\nu} &= \big( 1+\kappa\varphi l_{5}l_{5}\big)^{-\frac{2}{3}} \big(\eta_{\mu\nu} + \kappa\varphi\, l_{\mu}l_{\nu} \big)\,,
  \\
  g^{\mu\nu} &= \big( 1+\kappa\varphi l_{5}l_{5}\big)^{\frac{2}{3}} \Big(\ \eta^{\mu\nu} - \frac{\kappa\varphi}{1+\kappa\varphi l_{5}l_{5}} l^{\mu}l^{\nu}  \Big)\,,
  \\
  |g| &=  \big( 1+\kappa\varphi l_{5}l_{5}\big)^{-\frac{5}{3}} \,,
  \\
  v_{\mu} &= -(1+ \kappa\varphi l_{5}l_{5})^{\frac{1}{6}} \kappa\varphi l_{\mu} l_{5} \,, \qquad g^{\mu\nu}v_{\nu} = -(1+ \kappa\varphi l_{5}l_{5})^{-\frac{1}{6}} \kappa\varphi l^{\mu} l_{5} \,, 
  \\
  C_{\mu\nu\rho} &= \epsilon_{\mu\nu\rho\sigma}\frac{\kappa\varphi}{1+\varphi l_{5}l_{5}} l_{5} l^{\sigma}\,.
\end{aligned}\label{5d_1l}
\end{equation}
Again, this is consistent with the $\mathbf{10}$-representation result in \eqref{KS_components}.

\subsection{Equations of motion}\label{section:B.2}
The equations of motion for $m_{mn}$ in $\mathbf{5}$ representation is given by the generalised curvature tensor \cite{Park:2013gaj,Park:2014una}
\begin{equation}
\begin{aligned}
  \delta S = \int_{\Sigma} \mathrm{d}^{10}X \sqrt{|G|}~\delta m^{mn} \tilde{\mathcal{R}}_{m n}
\end{aligned}\label{}
\end{equation}
where 
\begin{equation}
\begin{aligned}
  \tilde{\mathcal{R}}_{m n} &= -\frac{1}{2\sqrt{G}}\partial_{i}\Big(\sqrt{G} G^{ij} \partial_{j} m_{mn}\Big) -\frac{1}{6\sqrt{G}} \partial_{i}\Big(\sqrt{G} G^{ij} m_{mn} m^{pq}\partial_{j}m_{pq}\Big)
  \\
  &\quad-\frac{1}{4} m^{pq} \partial_{mp} m_{r s} \partial_{nq}m^{rs} -\frac{1}{2} m^{pq} {} \partial_{(m|r} m_{ps|} \partial_{n) q} m^{r s} +m^{pq}  \partial_{(m|q|} m_{n)s} \partial_{p r} m^{rs} 
  \\
  &\quad + \frac{1}{2}m^{pq}  \partial_{p r} m^{rs} \partial_{qs} m_{mn} -m^{pq} m^{rs}  \partial_{(m|p|} \partial_{n)r} m_{q s} +m^{pq} m^{rs} \partial_{pr} \partial_{(m|q|} m_{n)s}  
  \\
  &\quad +\frac{1}{4} m^{pq} m^{rs} \partial_{pr} \partial_{qs} m_{mn} -m^{pq}\partial_{p r} m^{rs} \partial_{(m|s|} m_{n)q} - m^{pq} \partial_{(m|r|} m^{rs} \partial_{n) p} m_{qs}
  \\
  &\quad -\frac{1}{4}m^{pq} m^{rs} m^{tu} \partial_{pr} m_{mt} \partial_{qs} m_{nu} +\frac{1}{2}m^{pq} m^{rs} m^{tu} \partial_{pr} m_{(m|t} \partial_{qu|} m_{n)s}
  \\
  &\quad +\frac{1}{2} m^{pq}\Big(\partial_{mp}\partial_{nq} \ln|G| -\frac{1}{2}\partial_{mp} G^{ij}\partial_{nq}G_{ij}-\frac{1}{4} m^{rs}\partial_{pr}m_{mn}\partial_{qs}\ln|G|
  \\
  &\qquad\qquad\quad -m^{rs}\partial_{(m|p|} m_{n)r} \partial_{qs}\ln|G|\Big) +\frac{1}{2}\partial_{(m|p|}m^{pq}\partial_{n)q}\ln|G|
\end{aligned}\label{Rtom}
\end{equation}

Now we consider the full contraction between $l^{m}$ and $\tilde{\mathcal{R}}_{mn}$ in order to construct  constraint on the null vector such as the geodesic condition in the KS ansatz for GR,
\begin{equation}
\begin{aligned}
    l^{m} l^{n} \tilde{\mathcal{R}}_{mn} &= -2\varphi v_{[m n]} v^{[mn]} + \varphi v^{m}{}_{m} v^{n}{}_{n} -\frac{1}{2}l^{m} l^{n} \Big(\partial_{m}{}^{p} \partial_{n p} \ln|G| -\frac{1}{2} \partial_{m}{}^{p} G_{ij}\partial_{np} G^{ij}\Big) = 0\,,
\end{aligned}\label{onshell1}
\end{equation}
where $v_{mn} = l^{p}\partial_{pm} l_{n}$. Since the last term is proportional to the zeroth-order of the equations of motion in \eqref{eom_5d}, $R^{(0)}{}_{mn}$, and we can ignore the term in on-shell backgrounds of $G_{ij}$. Then we have the following relation on $v_{mn}$
\begin{equation}
\begin{aligned}
  -2 v_{[m n]} v^{[mn]} +  v^{m}{}_{m} v^{n}{}_{n} =0	
\end{aligned}\label{}
\end{equation}
We further require that $v_{mn}$ is traceless, $v^{m}{}_{m} = 0$, and then the antisymmetric part of $v_{mn}$ becomes null,
\begin{equation}
  v_{[mn]} v^{[mn]}= 0 \,.
\label{onshell2}\end{equation}

Now we expand the equations of motion by substituting the KS ansatz for the small metric \eqref{gKS} into $\tilde{\mathcal{R}}_{mn}$ \eqref{Rtom}. One can show that considering $(m^{-1})^{mp} \tilde{\mathcal{R}}_{pn}$ is simpler than using $ \tilde{\mathcal{R}}_{pn}$, because the expansion of the former terminates at the second order, but the later ends at the cubic order
\begin{equation}
  m_{0mp}\big(m^{-1}\big)^{pq} \tilde{\mathcal{R}}_{qn} = \mathcal{R}^{(0)}{}_{mn} + \kappa \mathcal{R}^{(1)}{}_{mn} +\kappa^{2} \mathcal{R}^{(2)}{}_{mn}
\label{5d_eom}\end{equation}
where 
\begin{equation}
\begin{aligned}
  \mathcal{R}^{(0)}{}_{mn} &= \partial_{m}{}^{p}\partial_{np} \ln \sqrt{|G|} -\frac{1}{4}\partial_{m}{}^{p} G^{ij} \partial_{np} G_{ij}\,,
  \\
  \mathcal{R}^{(1)}{}_{mn} &= \varphi l_{m}l^{p}\tilde{\mathcal{R}}^{(0)}{}_{pn} +\frac{1}{\sqrt{G}} \partial^{pq}\Big[\sqrt{G} \partial_{(m|p|} \big(\varphi l_{n)} l_{q}\big)\Big] + \frac{1}{4\sqrt{G}}\partial^{pq}\Big[\sqrt{G}\partial_{pq} \big(\varphi l_{m} l_{n}\big)\Big]
  \\
  &\quad -\frac{1}{\sqrt{G}} \partial_{(m}{}^{p}\Big[\sqrt{G} \partial_{n)}{}^{q}\big(\varphi l_{p}l_{q} \big)\Big] -\frac{\varphi}{2} l_{p}l_{q}\Big[ \partial_{m}{}^{p}\partial_{n}{}^{q} \ln |G| + \frac{1}{2} \partial_{m}{}^{p} G^{ij} \partial_{n}{}^{q}G_{ij} \Big]
  \\
  &\quad -\frac{1}{2\sqrt{G}}\partial_{i}\Big(\sqrt{G} G^{ij} \partial_{j} \big(\varphi l_{m} l_{n}\big)\Big)\,,
  \\
  \mathcal{R}^{(2)}{}_{mn} &= -\frac{3}{2\sqrt{G}} \partial^{pq} \Big[\sqrt{G} \varphi^{2} l_{n} l_{[m} v_{pq]} \Big] -\varphi^{2}v_{[mp]}v_{n}{}^{p} -\varphi^{2} l_{m}\partial_{np}l_{q}v^{[pq]}\,.
\end{aligned}\label{eom_5d}
\end{equation}
Note that the field equations are not linear due to the presence of $\tilde{\mathcal{R}}^{(2)}$. However, we choose $G_{ij}$ to satisfy the zeroth order equation which is the same equation in $\mathbf{10}$-representation \eqref{zeroth_eom} and require a similar constraint on $v_{[mm']}$ as $V_{mm'}$ in $\mathbf{10}$-representation \eqref{10_on-shell_constraint}
\begin{equation}
  v_{[mn]} = l_{[m}q_{n]}\,, \qquad \mathrm{where}~ q_{m} l^{m} = 0\,.
\label{onshell3}\end{equation}
This is the on-shell constraint for $\mathbf{5}$-version and makes $\mathcal{R}^{(2)}$ vanish by itself. Then the remaining equation is the linear equation
\begin{equation}
  \mathcal{R}^{(1)}{}_{mn} = 0\,.
\label{}\end{equation}
Thus it is the same structure that vanishing $\mathcal{R}^{(2)}$ by the on-shell constraint as $\mathbf{10}$-representation.

For introducing the KS double copy, we now decompose $\mathbf{5}$ vector indices $m = \{\mu,5\}$ the field equation as
\begin{equation}
  \mathcal{R}_{mn} = \Big\{\mathcal{R}_{\mu\nu}\,,\mathcal{R}_{\mu5}\,,\mathcal{R}_{55}\Big\}
\label{}\end{equation}
where
\begin{equation}
\begin{aligned}
  \mathcal{R}_{\mu\nu} &= - \frac{1}{2\sqrt{|G|}} \Big(\partial^{\rho} \big( \sqrt{|G|} \partial_{\rho} ( \varphi l_{\mu} l_{\nu} ) \big) -2\partial^{\rho} \big( \sqrt{|G|} \partial_{(\mu} ( \varphi l_{\nu)}l_{\rho}) \big) +2\partial_{(\mu} \big( \sqrt{|G|} \partial_{\nu)} ( \varphi l_{5} l_{5} )\big) \Big)
  \\
  &\quad - \frac{1}{2 \sqrt{|G|} } \partial_{i} \Big( \sqrt{|G|} G^{i j} \partial_{j} (\varphi l_{\mu} l_{\nu} ) \Big)
  \\
  \mathcal{R}_{\mu5} &= -\frac{1}{2 \sqrt{|G|}} \Bigg( \partial_{\mu} \Big( \sqrt{|G|} \partial_{\rho} ( \varphi l^{\rho} l_{5}) \Big) + \partial_{i} \Big( \sqrt{|G|} G^{i j} \partial_{j} (\varphi l_{\mu} l_{5} ) \Big)\Bigg)
  \\
  \mathcal{R}_{55} &= \frac{1}{2 \sqrt{|G|}}\Bigg( \partial^{\rho} \Big(\sqrt{|G|} \partial_{\rho} ( \varphi l_{5} l_{5}) \Big)
  -2 \partial_{\rho} \Big(\sqrt{|G|} \partial_{\sigma} ( \varphi l^{\rho} l^{\sigma} ) \Big) -\partial_{i}\Big(\sqrt{G} G^{ij} \partial_{j} \big(\varphi l_{5} l_{5}\big)\Big)\Bigg)
\end{aligned}\label{}
\end{equation}
%

\section{Useful identities\label{sec:C}}
\begin{equation}
\begin{aligned}
  g_{\mu_{1} \mu_{2} , \nu_{1} \nu_{2}} &= (g_{\mu_{1} \nu_{1}}g_{\mu_{2} \nu_{2}}-g_{\mu_{1} \nu_{2}}g_{\mu_{2} \nu_{1}}) \\
  g^{\mu_{1} \mu_{2} , \nu_{1} \nu_{2}} &= (g^{\mu_{1} \nu_{1}}g^{\mu_{2} \nu_{2}}-g^{\mu_{1} \nu_{2}}g^{\mu_{2} \nu_{1}}) \\
  g_{\mu_{1} \mu_{2} , \rho_{1} \rho_{2}} g^{\rho_{1} \rho_{2} , \nu_{1} \nu_{2}} &= 4 \delta_{[\mu_{1}}^{\nu_{1}} \delta_{\mu_{2}]}^{\nu_{2}} = 2(\delta_{\mu_{1}}^{\nu_{1}} \delta_{\mu_{2}}^{\nu_{2}} - \delta_{\mu_{2}}^{\nu_{1}} \delta_{\mu_{1}}^{\nu_{2}})	
\end{aligned}\label{}
\end{equation}
Our choice of convention of the totally antisymmetric tensor density $\epsilon$ and the epsilon tensor $\varepsilon$ are 
\begin{equation}
\begin{aligned}
  \epsilon^{0 1 2 3}&=1, \qquad \varepsilon^{\mu \nu \rho \sigma}=|g|^{-\frac{1}{2}} \epsilon^{\mu \nu \rho \sigma} \,,
  \\
  \epsilon_{0 1 2 3}&=1, \qquad \varepsilon_{\mu \nu \rho \sigma}=|g|^{\frac{1}{2}} \epsilon_{\mu \nu \rho \sigma}\,, 
  \\
  \varepsilon^{\mu_{1} \cdots \mu_{4}}&= (-1)^{t}g^{\mu_{1}\mu_{1}'}\cdots g^{\mu_{4}\mu'_{4}}\varepsilon_{\mu'_{1} \cdots \mu'_{4}}	\,,
\end{aligned}\label{Levi-Civita}
\end{equation}
where the superscript $t$ denotes the number of time components.
Identities for the tensor density:
\begin{equation}
\begin{aligned}
  \epsilon^{\mu_{1} \nu_{1} \rho_{1} \sigma_{1}}\epsilon_{\mu_{2} \nu_{2} \rho_{2} \sigma_{2}} &= 4! \delta_{[\mu_{2}}{}^{	\mu_{1}}\delta_{\nu_{2}}{}^{\nu_{1}} \delta_{\rho_{2}}{}^{\rho_{1}} \delta_{\sigma_{2}]}{}^{\sigma_{1}} 
  \\
  \epsilon^{\mu_{1} \nu_{1} \rho_{1} \sigma_{1}} \epsilon^{\mu_{2} \nu_{2} \rho_{2} \sigma_{2}} &= -4! |g|g^{\mu_{2} \tau} g^{\nu_{2} \kappa} g^{\rho_{2} \lambda} g^{\sigma_{2} \alpha}\delta_{[\tau}{}^{\mu_{1}}\delta_{\kappa}{}^{\nu_{1}} \delta_{\lambda}{}^{\rho_{1}} \delta_{\alpha]}{}^{\sigma_{1}} 
  \\
  \epsilon_{\mu_{1} \nu_{1} \rho_{1} \sigma_{1}} \epsilon_{\mu_{2} \nu_{2} \rho_{2} \sigma_{2}} & = -4! |g|^{-1} g_{\mu_{1} \tau} g_{\nu_{1} \kappa} g_{\rho_{1} \lambda} g_{\sigma_{1} \alpha}\delta_{[\mu_{2}}{}^{\tau}\delta_{\nu_{2}}{}^{\kappa} \delta_{\rho_{2}}{}^{\lambda} \delta_{\sigma_{2}]}{}^{\alpha}
\end{aligned}\label{}
\end{equation}
%


\bibliography{SL5KS}

\end{document}